\documentclass[11 pt,a4paper]{article}
\usepackage{amsmath}
\usepackage[pdftex]{graphicx}
\usepackage{epsfig}
\usepackage{epstopdf}
\graphicspath{{pictures/}}
\usepackage{geometry}
\usepackage{cite}
\usepackage{amssymb}
\usepackage{amsfonts}
\usepackage{amsthm}
\DeclareUnicodeCharacter{200F}{}
\usepackage{hyperref}
\makeatletter
\newcommand{\row}[1]%
{\mathord{\buildrel{\lower3pt%
\hbox{$\scriptscriptstyle\rightarrow$}}\over #1}}

\newcommand{\dyadic}[1]{\mathord{\dyadic@rrow{#1}}}
\newcommand{\dyadic@rrow}[1]{
\begin{picture}(12,12)(-1,0)
\put(-3,12){\makebox(0,0)[t]{$\scriptscriptstyle\downarrow$}}
\put(-3,13){\makebox(0,0)[l]{$\scriptscriptstyle\longrightarrow$}}
\put(5,0){\makebox(0,0)[b]{$#1$}}
\end{picture}
}

\date{}
\begin{document}
	\title{{\bf  Generating multi-hops entangled network via spin Dipolar  interaction
	}}
	\author{A. R. Mohammed $ ^{1} $\footnote {E-mail a.radwan@azhar.edu.eg}~~T. M. El-Shahat $ ^{1} $\footnote {E-mail el\_shahat@yahoo.com}~~and N. Metwally $ ^{2,3} $\footnote {E-mail nmetwally@gmail.com}\\
		$^{1}$Math. Dept., Faculty of Science, Al-Azhar University, Assiut 71524, Egypt.\\
		$^{2}$Math. Dept., Faculty of Science, Aswan University, Aswan 81528, Egypt\\.
$^{3}$Department of mathematics, College of Science, Bahrain University,
Bahrain
	}
	\maketitle
	{\large \centerline{{\bf Abstract}}}
The possibility of generating  a multi-hops network  between different entangled nodes (qubits) via spin Dipolar interaction is examined. The negativity, tangle and the non-local coherent advantage are used as quantifiers of the generated  quantum correlations. The phenomena of the sudden death/birth is displayed for the entangled two nodes, while the sudden changes phenomena (increasing/ decreasing) is depicted for all  entangled three nodes. The amount of correlations between the different nodes depend on the initial network settings, where the largest amount is predicted if the network is initially  conducted via maximum entangled nodes.
The generated quantum correlations between each three nodes  are more robust than those generated between two nodes.  For the generated entangled two nodes, the direction of the interaction and its strength  have  a remarkable effect on the correlation behavior, while they  has a  slightly effect on the correlation of the three nodes.

	\vspace{1 mm}

	\section{Introduction}
Quantum networks  play  some important roles  in  the quantum information processing tasks, such as distributed quantum computations,  quantum teleportation and quantum cryptography \cite{doplar1,doplar2}. Design network  protocols depends on the efficiency of the  connections between the nodes, the amount of correlations between them and on the property of the resources  to connect and  transfer quantum information between each  nodes of the network.
\\
There are some applications are  implemented via  these networks. For example,  Zueco et al. studied the routing of quantum information in qubit chains \cite{doplar3}. The possibility of using  the optical photons as carriers of information between fixed trapped atomic  quantum memories  to generate  entangled quantum networks is examined by Duan and Monroe \cite{doplar4}. Chudzicki and Strauch \cite{doplar5} discussed  the routing
of quantum information, where they  showed perfect parallel state transfer is possible for certain networks of harmonic oscillator modes. Metwally \cite{doplar6} has generated a wireless quantum network between multi-hops by using logic quantum gates. Generating entangled network via DzyaloshinskiiMoriya (DM)interaction is investigated in  \cite{doplar7}. It is shown that,  that entanglement may be enhanced under a magnetic field \cite{doplar8,doplar9,doplar10}. Castro et al. have used of a Dipolar spin thermal system as a noisy quantum channel to perform quantum teleportation \cite{doplar11}.

 In this contribution, we are motivated to   introduce a theoretical technique to
generate entangled network by using different types of  entangled states, where it is assumed that, a source supplies the users with pairs of $X$ states. As it is displayed in Fig.(1), it is allowed that the  second and third nodes (qubits) interact locally via Dipolar interaction \cite{doplar11}.  Due to this interaction, an entangled network is generated between the four nodes.
We assume that, the initial network is  conducted via maximum entangle two nodes ($MM$ type), Werner-Werner entangled nodes ($WW$ type), or by using maximum-Werner entangled nodes ($MW$ type). The amount of quantum correlations that generated between the network's nodes are quantified by the negativity, non-local coherent advantage, and the tangle.  The effect of the interaction strength and its direction on the behavior of the generated quantum correlation is discussed.

The paper is organized as follows. In Sec.(\ref{Dip2}), a model  of the suggested  quantum network is described, where an explicit analytical forms of the final state between each two and three nodes are introduced. Moreover, we examine the possibility of  generating multi-hops network between either two or three nodes of the network. In Sec.(\ref{Dip3}), we recall, the negativity, non-local advantage and the tangle as  measures of the generated quantum correlations between  the  different  nodes. The  behavior of these quantifiers are investigated  numerically in Sec.(\ref{Dip4}).  Finally, our results are summarized in Sec.(\ref{Dip5}).

\section{The  system and its evaluation}\label{Dip2}
The spin Dipolar interaction arises from the magnetic field created by a magnetic moment of a spin-$\frac{1}{2}$. The Hamiltonian that describes the interaction is given by,

\begin{eqnarray}
\mathcal{H}=-\frac{1}{3}\row{S}_{1}^{T}\cdot \dyadic{T}\cdot\row{S}_{2},
\end{eqnarray}
where $ \mathcal{S}_{i}=\{\mathcal{S}_{i}^{x},\mathcal{S}_{i}^{y},\mathcal{S}_{i}^{z}\} $, $i=1,2$ are the spin operators, and the dyadic $\dyadic{T}$ is defined by $3\times 3$ matrix its diagonal elements are given by $\{\Delta-3\epsilon,\Delta+3\epsilon,-2\Delta\} $, with $ \Delta $ and $ \epsilon $ are the Dipolar coupling constants between the spins.  If $\Delta>0$ then, the interaction is switched on the $x-y$  plane, whereas if $ \Delta<0 $, the spin is directed along the z-axis. In the computational basis set $\{00,01,10,11\}$, the  Hamiltonian (1) may be written as,
\begin{eqnarray}
\mathcal{H}=\left(
\begin{array}{cccc}
\frac{\Delta }{6} & 0 & 0 & \frac{\epsilon }{2} \\
0 & -\frac{\Delta }{6} & -\frac{\Delta }{6} & 0 \\
0 & -\frac{\Delta }{6} & -\frac{\Delta }{6} & 0 \\
\frac{\epsilon }{2} & 0 & 0 & \frac{\Delta }{6} \\
\end{array}
\right),
\end{eqnarray}
Let us assume that,  Alice and Bob share an entangled hop consists  of four nodes,  where each pair of these nodes may be prepared in  a singlet or Werner states. It is allowed that, the terminals of each two nodes interact via Dipolar interaction. Due to this interaction one may generated entangled network either between two singlet nodes ($MM $network), or between two Werner nodes state ($WW$ network) or between singlet- Werner ($MW$ network). This suggested network is displayed in Fig.(1a). The mathematical description of this network is as  follows:
 Alice and Bob share  two pairs of the $X$ state as,
 \begin{eqnarray}
\rho_{1234}(0)=\rho_{12}\otimes\rho_{34},
\end{eqnarray}
where $ \rho_{12} $ and $ \rho_{34}$ are defined as,
\begin{eqnarray}
\begin{aligned}
\rho_{12}&=\frac{1}{4}(1+a_{1}\sigma_{x}^{(1)}\tau_{x}^{(2)}+b_{1}\sigma_{y}^{(1)}\tau_{y}^{(2)}+c_{1}\sigma_{z}^{(1)}\tau_{z}^{(2)}),\\
\rho_{34}&=\frac{1}{4}(1+a_{2}\sigma_{x}^{(3)}\tau_{x}^{(4)}+b_{2}\sigma_{y}^{(3)}\tau_{y}^{(4)}+c_{2}\sigma_{z}^{(3)}\tau_{z}^{(4)}),
\end{aligned}	
\end{eqnarray}
where $\sigma_i, \tau_i, i=1..3$ represent the Pauli-operators for the  second and third nodes, respectively.
The time evolution of the initial network is given by,
\begin{eqnarray}
\rho_{1234}(t)=\mathcal{U}_{23}\rho_{1234}(0)\mathcal{U}^{\dagger}_{23}
\end{eqnarray}
where $\mathcal{U}_{23} = e^{-i\mathcal{H}t}$ represents the
unitary operator between the second and third nodes. In the computational basis, this unitary may be written as,
\begin{eqnarray}
\begin{aligned}
\mathcal{U}_{23}(t)&=(r_{1}+r_{4})|00\rangle  \langle 00|+(r_{1}-r_{4})|01\rangle  \langle 01|+(r_{1}-r_{4})|10\rangle  \langle 10|+(r_{2}+r_{3})(|00\rangle  \langle 11|+|11\rangle  \langle 00|)\\
&+(r_{2}-r_{3})(|01\rangle  \langle 10|+|10\rangle  \langle 01|)+(r_{1}+r_{4})|11\rangle  \langle 11|,
\end{aligned}
\end{eqnarray}
where
\begin{eqnarray}
\begin{aligned}
r_1&=c_1 c_2 c_3-i s_1 s_2 s_3, \quad r_2=c_1 s_2 s_3-i c_2 c_3 s_1, \quad r_3=c_2 s_1 s_3-i c_1 c_3 s_2, \quad r_4=c_3 s_1 s_2-i c_1 c_2 s_3,\\
c_1&=\cos (\kappa_{x} \tau), \quad c_2=\cos (\kappa_{y} \tau), \quad c_3=\cos (\kappa_{z} \tau), \quad s_1=\sin (\kappa_{x} \tau),
 \quad s_2=\sin (\kappa_{y} \tau),\\
 s_3&=\sin (\kappa_{z} \tau), \quad
\kappa_{x}=1-3 \tilde{\epsilon}, \quad \kappa_{y}=1+3 \tilde{\epsilon}, \quad \kappa_{z}=-2, \mbox{where}\quad \tilde\epsilon=\frac{\epsilon}{\Delta}.
\end{aligned}
\end{eqnarray}

\begin{figure}[!h]
	\begin{center}
		\includegraphics[width=0.8\textwidth, height=170px]{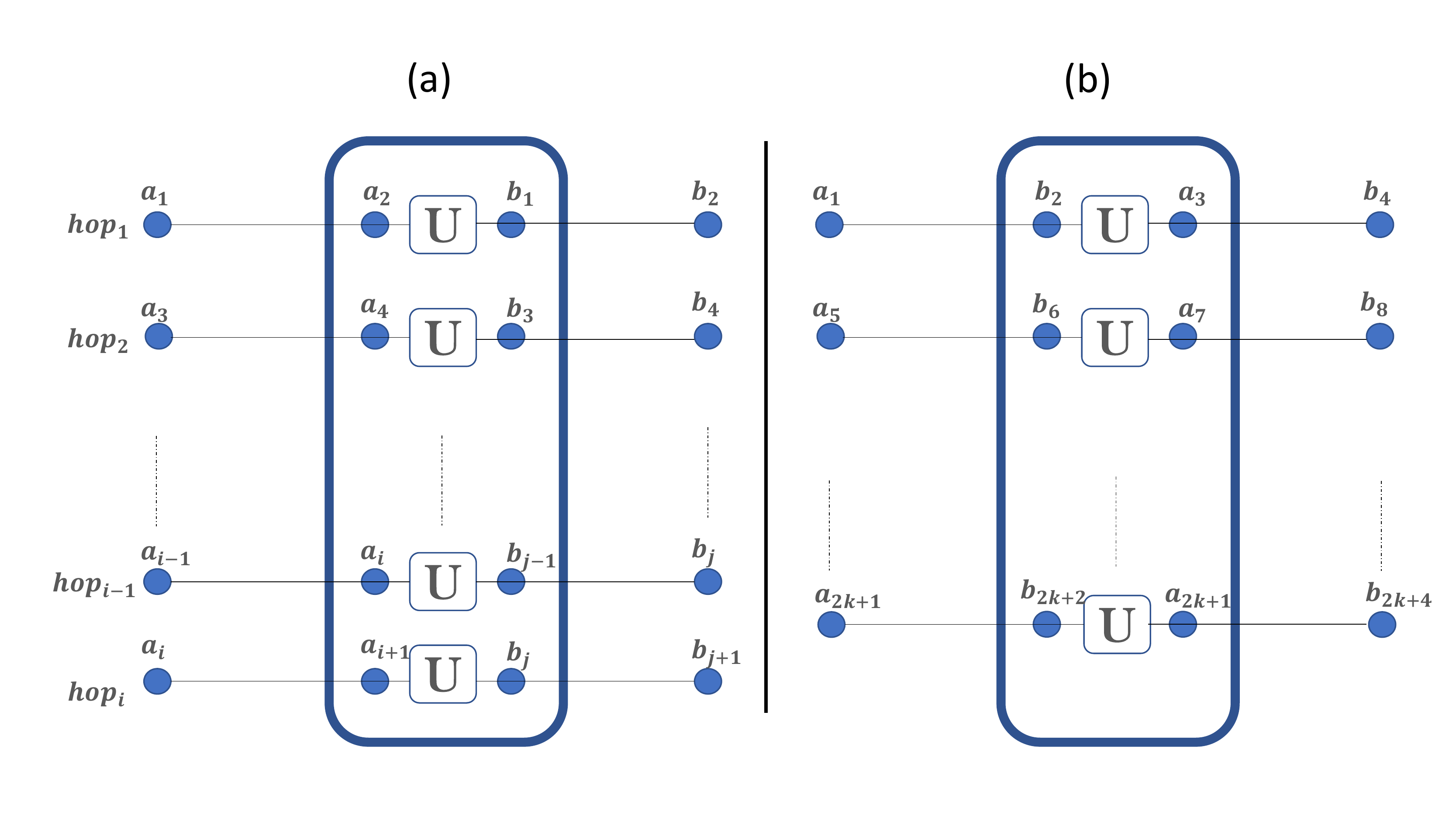}
	\end{center}
	\caption{\label{Dipolar1} This scheme describes the  suggested network (a) Alice and Bob  hold an $N$ entangled nodes $a_i,b_i, i=1...N$, respectively. The even nodes of Alice and the odd nodes of Bob  interacts locally via Dipolar interaction. Due to this interaction multi-hopes are generated between each four nodes,  (b) Shows the possibility of extending the network, where the  terminal of each hops interacts locally via Dipolar interaction  to generate  entangled hops of 8 nodes and so on.  }
\end{figure}
\subsection{Two entangled nodes}
To discuss the  entanglement robustness of the channel $\rho_{12}(t)$ between the first and the second nodes, one  traces out the other nodes from the final total state $\rho_{1234}(t)$, namely $\rho_{12}(t)=Tr_{34}\{\rho _{1234}(t)\}$. The final network state  $\rho_{12}(t)$ is defined by  a $4\times 4$ matrix. Its non-zero elements, the final state between the first and the second nodes are given by,

\begin{eqnarray}
\begin{aligned}
\rho _{11}&=2 B_1 (A_1 r_1 r_1^\ast+A_2 r_2 r_2^\ast+A_2 r_3 r_3^\ast+A_1 r_4 r_4^\ast),
\rho _{14}=2 B _1 (A_3 r_1 r_1^\ast+A_4 r_2 r_2^\ast-A_4 r_3 r_3^\ast-A_3 r_4 r_4^\ast),\\
\rho _{22}&=2 B _1 (A_2 r_1 r_1^\ast+A_1 r_2 r_2^\ast+A_1 r_3 r_3^\ast+A_2 r_4 r_4^\ast),
\rho _{23}=2 B _1 (A_4 r_1 r_1^\ast+A_3 r_2 r_2^\ast-A_3 r_3 r_3^\ast-A_4 r_4 r_4^\ast),\\
\rho _{32}&=\rho _{23},\quad\rho _{33}=\rho _{22},\quad\rho _{41}=\rho _{14},\quad\rho _{44}=\rho _{11},\quad\beta _1=B_1+B_2,\quad\beta _2=A_1+A_2,
\end{aligned}
\end{eqnarray}
where
\begin{eqnarray*}
A_1&=&\frac{1}{4} (c_1+1),\quad A_2=\frac{1}{4} (1-c_1),\quad A_3=\frac{1}{4} (a_1-b_1),\quad A_3=\frac{1}{4} (a_1+b_1),
\nonumber\\
B_1&=&\frac{1}{4} (c_2+1),\quad B_2=\frac{1}{4} (1-c_2),\quad B_3=\frac{1}{4} (a_2-b_2),\quad B_3=\frac{1}{4} (a_2+b_2).
\end{eqnarray*}

Due to the interaction, an entangled state is generated between the first and the fourth nodes, $ \rho_{14}=Tr_{23}( \rho _{1234}(t)) $ which may  be described by the following non zero matrix  of size $ 4\times 4$:

\begin{eqnarray}
\begin{aligned}
\rho _{11}&=A_1 (B_2 \gamma _1 \gamma _1^\ast+B_1 \gamma _2 \gamma _2^\ast+B_2 \gamma _3 \gamma _3^\ast+B_1 \gamma _4 \gamma _4^\ast)
+A_2 (B_1 \gamma _1 \gamma _1^\ast+B_2 \gamma _2 \gamma _2^\ast+B_1 \gamma _3 \gamma _3^\ast+B_2 \gamma _4 \gamma _4^\ast),\\
\rho _{14}&=A_3 (B_4 \gamma _3 \gamma _1^\ast+B_3 \gamma _4 \gamma _2^\ast+B_4 \gamma _1 \gamma _3^\ast+B_3 \gamma _2 \gamma _4^\ast)
+A_4 (B_3 \gamma _3 \gamma _1^\ast+B_4 \gamma _4 \gamma _2^\ast+B_3 \gamma _1 \gamma _3^\ast+B_4 \gamma _2 \gamma _4^\ast),\\
\rho _{22}&=A_2 (B_2 \gamma _1 \gamma _1^\ast+B_1 \gamma _2 \gamma _2^\ast+B_2 \gamma _3 \gamma _3^\ast+B_1 \gamma _4 \gamma _4^\ast)
+A_1 (B_1 \gamma _1 \gamma _1^\ast+B_2 \gamma _2 \gamma _2^\ast+B_1 \gamma _3 \gamma _3^\ast+B_2 \gamma _4 \gamma _4^\ast),\\
\rho _{23}&=A_4 (B_4 \gamma _3 \gamma _1^\ast+B_3 \gamma _4 \gamma _2^\ast+B_4 \gamma _1 \gamma _3^\ast+B_3 \gamma _2 \gamma _4^\ast)
+A_3 (B_3 \gamma _3 \gamma _1^\ast+B_4 \gamma _4 \gamma _2^\ast+B_3 \gamma _1 \gamma _3^\ast+B_4 \gamma _2 \gamma _4^\ast),\\
\rho _{32}&=\rho _{23},\quad\rho _{33}=\rho _{22},\quad\rho _{41}=\rho _{14},\quad\rho _{44}=\rho _{11},\quad\gamma _1=r_2-r_3,\quad\gamma _2=r_2+r_3,\quad\gamma _3=r_1-r_4,\\
\gamma _4&=r_1+r_4.
\end{aligned}
\end{eqnarray}
As a direct interaction between the second and third nodes,  an entangled state $\rho_{23}=Tr_{14}\{ \rho _{1234}(t)\} $ is generated. Its non-zero elements are given  by,
\begin{eqnarray}
\begin{aligned}
\rho _{11}&=\beta _2 \beta _1 (\gamma _2 \gamma _2^\ast+\gamma _4 \gamma _4^\ast), \quad
\rho _{14}=\beta _2 \beta _1 (\gamma _4 \gamma _2^\ast+\gamma _2 \gamma _4^\ast),\\
\rho _{22}&=\beta _2 \beta _1 (\gamma _1 \gamma _1^\ast+\gamma _3 \gamma _3^\ast),\quad
\rho _{23}=\beta _2 \beta _1 (\gamma _3 \gamma _1^\ast+\gamma _1 \gamma _3^\ast),\\
\rho _{32}&=\rho _{23},\quad\rho _{33}=\rho _{22},\quad\rho _{41}=\rho _{14},\quad\rho _{44}=\rho _{11}.
\end{aligned}
\end{eqnarray}

\subsection{Three entangled nodes}
In this subsection, we investigate the amount of correlations that may be generated between three different nodes via the spin Dipolar interaction. There are three possibilities of generated $3$-nodes channels, either $\rho_{123}$, $\rho_{234}$, or $\rho_{124}$.  Each  channel is described by a matrix of size $8\times 8$ elements. The three nodes channel $\rho_{123}(t)$  is given by,
\begin{eqnarray}
\begin{aligned}
\rho_{123}(t)&=\beta _1 (A_2 \gamma _2 \gamma _2^\ast+A_1 \gamma _4 \gamma _4^\ast)(|000\rangle  \langle 000|+|111\rangle  \langle 111|)
+\beta _1 (A_2 \gamma _4 \gamma _2^\ast+A_1 \gamma _2 \gamma _4^\ast)(|011\rangle  \langle 000|+|100\rangle  \langle 111|)\\
&+\beta _1 (A_4 \gamma _3 \gamma _2^\ast+A_3 \gamma _1 \gamma _4^\ast)(|101\rangle  \langle 000|+|010\rangle  \langle 111|)
+\beta _1 (A_4 \gamma _1 \gamma _2^\ast+A_3 \gamma _3 \gamma _4^\ast)(|110\rangle  \langle 000|+|001\rangle  \langle 111|)\\
&+\beta _1 (A_2 \gamma _1 \gamma _1^\ast+A_1 \gamma _3 \gamma _3^\ast)(|001\rangle  \langle 001|+|110\rangle  \langle 110|)
+\beta _1 (A_2 \gamma _3 \gamma _1^\ast+A_1 \gamma _1 \gamma _3^\ast)(|010\rangle  \langle 001|+|101\rangle  \langle 110|)\\
&+\beta _1 (A_4 \gamma _4 \gamma _1^\ast+A_3 \gamma _2 \gamma _3^\ast)(|100\rangle  \langle 001|+|011\rangle  \langle 110|)
+\beta _1 (A_4 \gamma _2 \gamma _1^\ast+A_3 \gamma _4 \gamma _3^\ast)(|111\rangle  \langle 001|+|000\rangle  \langle 110|)\\
&+\beta _1 (A_1 \gamma _3 \gamma _1^\ast+A_2 \gamma _1 \gamma _3^\ast)(|001\rangle  \langle 010|+|110\rangle  \langle 101|)
+\beta _1 (A_1 \gamma _1 \gamma _1^\ast+A_2 \gamma _3 \gamma _3^\ast)(|010\rangle  \langle 010|+|101\rangle  \langle 101|)\\
&+\beta _1 (A_3 \gamma _2 \gamma _1^\ast+A_4 \gamma _4 \gamma _3^\ast)(|100\rangle  \langle 010|+|011\rangle  \langle 101|)
+\beta _1 (A_3 \gamma _4 \gamma _1^\ast+A_4 \gamma _2 \gamma _3^\ast)(|111\rangle  \langle 010|+|000\rangle  \langle 101|)\\
&+\beta _1 (A_1 \gamma _4 \gamma _2^\ast+A_2 \gamma _2 \gamma _4^\ast)(|000\rangle  \langle 011|+|111\rangle  \langle 100|)
+\beta _1 (A_1 \gamma _2 \gamma _2^\ast+A_2 \gamma _4 \gamma _4^\ast)(|011\rangle  \langle 011|+|100\rangle  \langle 100|)\\
&+\beta _1 (A_3 \gamma _1 \gamma _2^\ast+A_4 \gamma _3 \gamma _4^\ast)(|101\rangle  \langle 011|+|010\rangle  \langle 100|)
+\beta _1 (A_3 \gamma _3 \gamma _2^\ast+A_4 \gamma _1 \gamma _4^\ast)(|110\rangle  \langle 011|+|001\rangle  \langle 100|).
\end{aligned}
\end{eqnarray}
Similarly the three nodes channel $\rho_{234}$ is  given by,
\begin{eqnarray}
\begin{aligned}
\rho_{(234)}(t)&=\beta _2 (B_2 \gamma _2 \gamma _2^\ast+B_1 \gamma _4 \gamma _4^\ast)(|000\rangle  \langle 000|+|111\rangle  \langle 111|)
+\beta _2 (B_4 \gamma _1 \gamma _2^\ast+B_3 \gamma _3 \gamma _4^\ast)(|011\rangle  \langle 000|+|100\rangle  \langle 111|)\\
&+\beta _2 (B_4 \gamma _3 \gamma _2^\ast+B_3 \gamma _1 \gamma _4^\ast)(|101\rangle  \langle 000|+|010\rangle  \langle 111|)
+\beta _2 (B_2 \gamma _4 \gamma _2^\ast+B_1 \gamma _2 \gamma _4^\ast)(|110\rangle  \langle 000|+|001\rangle  \langle 111|)\\
&+\beta _2 (B_1 \gamma _2 \gamma _2^\ast+B_2 \gamma _4 \gamma _4^\ast)(|001\rangle  \langle 001|+|110\rangle  \langle 110|)
+\beta _2 (B_3 \gamma _1 \gamma _2^\ast+B_4 \gamma _3 \gamma _4^\ast)(|010\rangle  \langle 001|+|101\rangle  \langle 110|)\\
&+\beta _2 (B_3 \gamma _3 \gamma _2^\ast+B_4 \gamma _1 \gamma _4^\ast)(|100\rangle  \langle 001|+|011\rangle  \langle 110|)
+\beta _2 (B_1 \gamma _4 \gamma _2^\ast+B_2 \gamma _2 \gamma _4^\ast)(|111\rangle  \langle 001|+|000\rangle  \langle 110|)\\
&+\beta _2 (B_3 \gamma _2 \gamma _1^\ast+B_4 \gamma _4 \gamma _3^\ast)(|001\rangle  \langle 010|+|110\rangle  \langle 101|)
+\beta _2 (B_1 \gamma _1 \gamma _1^\ast+B_2 \gamma _3 \gamma _3^\ast)(|010\rangle  \langle 010|+|101\rangle  \langle 101|)\\
&+\beta _2 (B_1 \gamma _3 \gamma _1^\ast+B_2 \gamma _1 \gamma _3^\ast)(|100\rangle  \langle 010|+|011\rangle  \langle 101|)
+\beta _2 (B_3 \gamma _4 \gamma _1^\ast+B_4 \gamma _2 \gamma _3^\ast)(|111\rangle  \langle 010|+|000\rangle  \langle 101|)\\
&+\beta _2 (B_4 \gamma _2 \gamma _1^\ast+B_3 \gamma _4 \gamma _3^\ast)(|000\rangle  \langle 011|+|111\rangle  \langle 100|)
+\beta _2 (B_2 \gamma _1 \gamma _1^\ast+B_1 \gamma _3 \gamma _3^\ast)(|011\rangle  \langle 011|+|100\rangle  \langle 100|)\\
&+\beta _2 (B_2 \gamma _3 \gamma _1^\ast+B_1 \gamma _1 \gamma _3^\ast)(|101\rangle  \langle 011|+|010\rangle  \langle 100|)
+\beta _2 (B_2 \gamma _3 \gamma _1^\ast+B_1 \gamma _1 \gamma _3^\ast)(|110\rangle  \langle 011|+|001\rangle  \langle 100|).
\end{aligned}
\end{eqnarray}
Finally, the channel that connect the first, second and the fourth nodes  is defined as,

\begin{eqnarray}
\begin{aligned}
\rho_{124}&=(A_2 B_1 \gamma _1 \gamma _1^\ast+A_2 B_2 \gamma _2 \gamma _2^\ast+A_1 B_2 \gamma _3 \gamma _3^\ast+A_1 B_1 \gamma _4 \gamma _4^\ast)|000\rangle  \langle 000|\\
&+(A_2 B_3 \gamma _4 \gamma _1^\ast+A_2 B_4 \gamma _3 \gamma _2^\ast+A_1 B_4 \gamma _2 \gamma _3^\ast+A_1 B_3 \gamma _1 \gamma _4^\ast)(|011\rangle  \langle 000|+|100\rangle  \langle 111|+|111\rangle  \langle 111|)\\
&+(A_4 B_3 \gamma _3 \gamma _1^\ast+A_4 B_4 \gamma _4 \gamma _2^\ast+A_3 B_4 \gamma _1 \gamma _3^\ast+A_3 B_3 \gamma _2 \gamma _4^\ast)(|101\rangle  \langle 000|+|010\rangle  \langle 111|)\\
&+(A_4 B_1 \gamma _2 \gamma _1^\ast+A_4 B_2 \gamma _1 \gamma _2^\ast+A_3 B_2 \gamma _4 \gamma _3^\ast+A_3 B_1 \gamma _3 \gamma _4^\ast)(|110\rangle  \langle 000|+|001\rangle  \langle 111|)\\
&+(A_2 B_2 \gamma _1 \gamma _1^\ast+A_2 B_1 \gamma _2 \gamma _2^\ast+A_1 B_1 \gamma _3 \gamma _3^\ast+A_1 B_2 \gamma _4 \gamma _4^\ast)(|001\rangle  \langle 001|+|110\rangle  \langle 110|)\\
&+(A_2 B_4 \gamma _4 \gamma _1^\ast+A_2 B_3 \gamma _3 \gamma _2^\ast+A_1 B_3 \gamma _2 \gamma _3^\ast+A_1 B_4 \gamma _1 \gamma _4^\ast)(|010\rangle  \langle 001|+|101\rangle  \langle 110|)\\
&+(A_4 B_4 \gamma _3 \gamma _1^\ast+A_4 B_3 \gamma _4 \gamma _2^\ast+A_3 B_3 \gamma _1 \gamma _3^\ast+A_3 B_4 \gamma _2 \gamma _4^\ast)(|100\rangle  \langle 001|+|011\rangle  \langle 110|)\\
&+(A_4 B_2 \gamma _2 \gamma _1^\ast+A_4 B_1 \gamma _1 \gamma _2^\ast+A_3 B_1 \gamma _4 \gamma _3^\ast+A_3 B_2 \gamma _3 \gamma _4^\ast)(|111\rangle  \langle 001|+|000\rangle  \langle 110|)\\
&+(A_1 B_4 \gamma _4 \gamma _1^\ast+A_1 B_3 \gamma _3 \gamma _2^\ast+A_2 B_3 \gamma _2 \gamma _3^\ast+A_2 B_4 \gamma _1 \gamma _4^\ast)(|001\rangle  \langle 010|+|110\rangle  \langle 101|)\\
&+(A_1 B_2 \gamma _1 \gamma _1^\ast+A_1 B_1 \gamma _2 \gamma _2^\ast+A_2 B_1 \gamma _3 \gamma _3^\ast+A_2 B_2 \gamma _4 \gamma _4^\ast)(|010\rangle  \langle 010|+|101\rangle  \langle 101|)\\
&+(A_3 B_2 \gamma _2 \gamma _1^\ast+A_3 B_1 \gamma _1 \gamma _2^\ast+A_4 B_1 \gamma _4 \gamma _3^\ast+A_4 B_2 \gamma _3 \gamma _4^\ast)(|100\rangle  \langle 010|+|011\rangle  \langle 101|)\\
&+(A_3 B_4 \gamma _3 \gamma _1^\ast+A_3 B_3 \gamma _4 \gamma _2^\ast+A_4 B_3 \gamma _1 \gamma _3^\ast+A_4 B_4 \gamma _2 \gamma _4^\ast)(|111\rangle  \langle 010|+|000\rangle  \langle 101|)\\
&+(A_1 B_3 \gamma _4 \gamma _1^\ast+A_1 B_4 \gamma _3 \gamma _2^\ast+A_2 B_4 \gamma _2 \gamma _3^\ast+A_2 B_3 \gamma _1 \gamma _4^\ast)(|000\rangle  \langle 011|+|111\rangle  \langle 100|)\\
&+(A_1 B_1 \gamma _1 \gamma _1^\ast+A_1 B_2 \gamma _2 \gamma _2^\ast+A_2 B_2 \gamma _3 \gamma _3^\ast+A_2 B_1 \gamma _4 \gamma _4^\ast)(|011\rangle  \langle 011|+|100\rangle  \langle 100|)\\
&+(A_3 B_1 \gamma _2 \gamma _1^\ast+A_3 B_2 \gamma _1 \gamma _2^\ast+A_4 B_2 \gamma _4 \gamma _3^\ast+A_4 B_1 \gamma _3 \gamma _4^\ast)(|101\rangle  \langle 011|+|010\rangle  \langle 100|)\\
&+(A_3 B_3 \gamma _3 \gamma _1^\ast+A_3 B_4 \gamma _4 \gamma _2^\ast+A_4 B_4 \gamma _1 \gamma _3^\ast+A_4 B_3 \gamma _2 \gamma _4^\ast)(|110\rangle  \langle 011|+|001\rangle  \langle 100|).
\end{aligned}
\end{eqnarray}

\subsection{Extension of Network}
It is assumed that, we have  two  networks each consists of four nodes. These two networks are connected via spin Dipolar interaction where  the  terminal node  of the first network interacts with the first node of the second network to generate an entangled hop consists of $8$ nodes. We continue to generate a large network by allowing these hops of the eighth nodes interact to generate a larger entangled hops and so on. In the following we consider only the channel between the terminals of this network, namely the channel $\rho_{18}$. The non-zero elements of the $4\times 4$ matrix are given by,

\begin{eqnarray}
\begin{aligned}
\rho _{11}&=\delta _3^2 ((\gamma _2 \gamma _2^\ast+\gamma _4 \gamma _4^\ast) M_{11}^2+(\gamma _1 \gamma _1^\ast+\gamma _3 \gamma _3^\ast) (M_{22}+M_{33}) M_{11}+(\gamma _2 \gamma _2^\ast+\gamma _4 \gamma _4^\ast) M_{22} M_{33}),\\
\rho _{14}&=\delta _3^2 ((\gamma _4 \gamma _2^\ast+\gamma _2 \gamma _4^\ast) M_{14}^2+(\gamma _3 \gamma _1^\ast+\gamma _1 \gamma _3^\ast) (M_{23}+M_{32}) M_{14}+(\gamma _4 \gamma _2^\ast+\gamma _2 \gamma _4^\ast) M_{23} M_{32}),\\
\rho _{22}&=\delta _3^2 (M_{11} (\gamma _1 \gamma _1^\ast M_{44}+\gamma _2 \gamma _2^\ast M_{22}+\gamma _3 \gamma _3^\ast M_{44}+\gamma _4 \gamma _4^\ast M_{22})+M_{22} (\gamma _1 \gamma _1^\ast M_{22}+\gamma _2 \gamma _2^\ast M_{44}\\
&+\gamma _3 \gamma _3^\ast M_{22}+\gamma _4 \gamma _4^\ast M_{44})),\\
\rho _{23}&=\delta _3^2 (M_{14} (\gamma _3 \gamma _1^\ast M_{41}+\gamma _4 \gamma _2^\ast M_{23}+\gamma _1 \gamma _3^\ast M_{41}+\gamma _2 \gamma _4^\ast M_{23})+M_{23} (\gamma _3 \gamma _1^\ast M_{23}+\gamma _4 \gamma _2^\ast M_{41}\\
&+\gamma _1 \gamma _3^\ast M_{23}+\gamma _2 \gamma _4^\ast M_{41})),\\
\rho _{32}&=\delta _3^2 (M_{14} (\gamma _3 \gamma _1^\ast M_{41}+\gamma _4 \gamma _2^\ast M_{32}+\gamma _1 \gamma _3^\ast M_{41}+\gamma _2 \gamma _4^\ast M_{32})+M_{32} (\gamma _3 \gamma _1^\ast M_{32}+\gamma _4 \gamma _2^\ast M_{41}\\
&+\gamma _1 \gamma _3^\ast M_{32}+\gamma _2 \gamma _4^\ast M_{41})),\\
\rho _{33}&=\delta _3^2 (M_{11} (\gamma _1 \gamma _1^\ast M_{44}+\gamma _2 \gamma _2^\ast M_{33}+\gamma _3 \gamma _3^\ast M_{44}+\gamma _4 \gamma _4^\ast M_{33})+M_{33} (\gamma _1 \gamma _1^\ast M_{33}+\gamma _2 \gamma _2^\ast M_{44}\\
&+\gamma _3 \gamma _3^\ast M_{33}+\gamma _4 \gamma _4^\ast M_{44})),\\
\rho _{41}&=\delta _3^2 (M_{23} (\gamma _3 \gamma _1^\ast M_{41}+\gamma _4 \gamma _2^\ast M_{32}+\gamma _1 \gamma _3^\ast M_{41}+\gamma _2 \gamma _4^\ast M_{32})+M_{41} (\gamma _3 \gamma _1^\ast M_{32}+\gamma _4 \gamma _2^\ast M_{41}\\
&+\gamma _1 \gamma _3^\ast M_{32}+\gamma _2 \gamma _4^\ast M_{41})),\\
\rho _{44}&=\delta _3^2 (M_{22} (\gamma _1 \gamma _1^\ast M_{44}+\gamma _2 \gamma _2^\ast M_{33}+\gamma _3 \gamma _3^\ast M_{44}+\gamma _4 \gamma _4^\ast M_{33})+M_{44} (\gamma _1 \gamma _1^\ast M_{33}+\gamma _2 \gamma _2^\ast M_{44}\\
&+\gamma _3 \gamma _3^\ast M_{33}+\gamma _4 \gamma _4^\ast M_{44})),
\end{aligned}
\end{eqnarray}
where,

\begin{eqnarray}
\delta _1&=&N_{11}+N_{22},\quad\delta _2=N_{33}+N_{44}, \delta_3=N_{11}+N_{22}+N_{33}+N_{44},
\nonumber\\
\delta _4&=&(\gamma _2 \gamma _2^\ast+\gamma _4 \gamma _4^\ast) M_{11}^2+M_{22} ((\gamma _1 \gamma _1^\ast+\gamma _3 \gamma _3^\ast) M_{22}+(\gamma _2 \gamma _2^\ast+\gamma _4 \gamma _4^\ast) (M_{33}+M_{44}))
\nonumber\\
&&+M_{11} ((\gamma _1 \gamma _1^\ast+\gamma _3 \gamma _3^\ast) (M_{33}+M_{44})+2 M_{22} (r_1 (r_1){}^*+r_2 (r_2){}^*+r_3 (r_3){}^*+r_4 (r_4){}^*)),
\nonumber\\
\delta _5&=&M_{11} (\gamma _1 \gamma _1^\ast M_{44}+\gamma _2 \gamma _2^\ast M_{33}+\gamma _3 \gamma _3^\ast M_{44}+\gamma _4 \gamma _4^\ast M_{33})+M_{22} (\gamma _1 \gamma _1^\ast M_{44}+\gamma _2 \gamma _2^\ast M_{33}
\nonumber\\
&&+\gamma _3 \gamma _3^\ast M_{44}+\gamma _4 \gamma _4^\ast M_{33})+(M_{33}+M_{44}) (\gamma _1 \gamma _1^\ast M_{33}+\gamma _2 \gamma _2^\ast M_{44}+\gamma _3 \gamma _3^\ast M_{33}+\gamma _4 \gamma _4^\ast M_{44}).
\end{eqnarray}
and,
\begin{eqnarray}
\begin{aligned}
M _{11}&=A_1 (B_2 \gamma _1 \gamma _1^\ast+B_1 \gamma _2 \gamma _2^\ast+B_2 \gamma _3 \gamma _3^\ast+B_1 \gamma _4 \gamma _4^\ast)+A_2 (B_1 \gamma _1 \gamma _1^\ast+B_2 \gamma _2 \gamma _2^\ast+B_1 \gamma _3 \gamma _3^\ast+B_2 \gamma _4 \gamma _4^\ast),\\
M _{14}&=A_3 (B_4 \gamma _3 \gamma _1^\ast+B_3 \gamma _4 \gamma _2^\ast+B_4 \gamma _1 \gamma _3^\ast+B_3 \gamma _2 \gamma _4^\ast)+A_4 (B_3 \gamma _3 \gamma _1^\ast+B_4 \gamma _4 \gamma _2^\ast+B_3 \gamma _1 \gamma _3^\ast+B_4 \gamma _2 \gamma _4^\ast),\\
M _{22}&=A_2 (B_2 \gamma _1 \gamma _1^\ast+B_1 \gamma _2 \gamma _2^\ast+B_2 \gamma _3 \gamma _3^\ast+B_1 \gamma _4 \gamma _4^\ast)+A_1 (B_1 \gamma _1 \gamma _1^\ast+B_2 \gamma _2 \gamma _2^\ast+B_1 \gamma _3 \gamma _3^\ast+B_2 \gamma _4 \gamma _4^\ast),\\
M _{23}&=A_4 (B_4 \gamma _3 \gamma _1^\ast+B_3 \gamma _4 \gamma _2^\ast+B_4 \gamma _1 \gamma _3^\ast+B_3 \gamma _2 \gamma _4^\ast)+A_3 (B_3 \gamma _3 \gamma _1^\ast+B_4 \gamma _4 \gamma _2^\ast+B_3 \gamma _1 \gamma _3^\ast+B_4 \gamma _2 \gamma _4^\ast),\\
M _{32}&=M _{23},\quad M _{33}=M _{22},\quad M _{41}=M _{14},\quad M _{44}=M _{11},\quad
N _{11}=\beta _2 \beta _1 (\gamma _2 \gamma _2^\ast+\gamma _4 \gamma _4^\ast),\\
N _{14}&=\beta _2 \beta _1 (\gamma _4 \gamma _2^\ast+\gamma _2 \gamma _4^\ast),\quad
N _{22}=\beta _2 \beta _1 (\gamma _1 \gamma _1^\ast+\gamma _3 \gamma _3^\ast),\quad
N _{23}=\beta _2 \beta _1 (\gamma _3 \gamma _1^\ast+\gamma _1 \gamma _3^\ast),\\
N _{32}&=N _{23},\quad N _{33}=N _{22},\quad N _{41}=N _{14},\quad N _{44}=N _{11}.
\end{aligned}
\end{eqnarray}
\section{Quantifying the correlations}\label{Dip3}
\begin{itemize}
\item{\it Two nodes entanglement:Negativity\\}
Negativity is considered as one of the most common entanglement  measure  of  a two-qubit system, where it satisfies all the conditions as a measure of entanglement \cite{cohe21}. However, for any quantum state $\varrho_{ab}$, the negativity  is defined by means of the eigenvalues of the partial transpose of the state $\varrho_{ab}$. Mathematically, it may be defined as,
\begin{equation}
\mathcal{N}_{eg}=2\sum_{i}|\lambda_i|,
\end{equation}
where $\lambda_i, i=1,2,3,4$ are the eigenvalues of the density operator $\varrho_{ab}^{T_b}$.

\item{\it Three nodes entanglement:Tangle\\}
To quantify  the amount of entanglement that may be generated between each three nodes($3$ qubits) is quantified by  the $ \pi $-tangle which proved to be a natural entanglement measure \cite{doplar12}. For a three-qubit (node) state $ \rho $, a monogamy inequality analogous to Coffman-Kundu-Wootters (CKW) inequality is satisfies  on the single  qubit  as the entanglement quantified by the negativity between A and B;  A and C, and, between $A$, $BC$ as following,
\begin{eqnarray}
\begin{aligned}
\mathcal{N}_{AB}^{2}+\mathcal{N}_{AC}^{2}\leq \mathcal{N}_{A(BC)}^{2}.
\end{aligned}
\end{eqnarray}
In a similar way, if one takes the different focus B and C, the following monogamy inequalities  are defined by,
\begin{eqnarray}
\begin{aligned}
&\mathcal{N}_{BA}^{2}+\mathcal{N}_{BC}^{2}\leq \mathcal{N}_{B(AC)}^{2},\\
&\mathcal{N}_{CA}^{2}+\mathcal{N}_{CB}^{2}\leq \mathcal{N}_{C(AB)}^{2},\\
\end{aligned}
\end{eqnarray}
The $ \pi $-tangle is defined in terms of the global negativities \cite{doplar13} as,
\begin{eqnarray}
\begin{aligned}
\mathcal{N}^{A}=\parallel \rho^{T_{A}}\parallel-1,\qquad\mathcal{N}^{B}=\parallel \rho^{T_{B}}\parallel-1,\qquad\mathcal{N}^{C}=\parallel \rho^{T_{C}}\parallel-1.
\end{aligned}
\end{eqnarray}
 The superscripts $ T_{A} $, $ T_{B} $, and $ T_{C} $ represent the partial transposes
of $ \rho $. Then, the $ \pi $-tangle is defined as,
\begin{eqnarray}
\begin{aligned}
\pi_{ABC}=\frac{1}{3}(\pi_{A}+\pi_{B}+\pi_{C}),
\end{aligned}
\end{eqnarray}
where
\begin{eqnarray}
\begin{aligned}
\pi_{A}&=\mathcal{N}_{A(BC)}^{2}-(\mathcal{N}_{(AB)}^{2}+\mathcal{N}_{(AC)}^{2}),\\
\pi_{B}&=\mathcal{N}_{B(AC)}^{2}-(\mathcal{N}_{(AB)}^{2}+\mathcal{N}_{(BC)}^{2}),\\
\pi_{C}&=\mathcal{N}_{C(AB)}^{2}-(\mathcal{N}_{(AC)}^{2}+\mathcal{N}_{(BC)}^{2}).\\
\end{aligned}
\end{eqnarray}
\item{Non-local advantage\\}
For a two-qubit state $\rho_{AB} $, based on local measurements on subsystem A and classical communication between the bipartite, the nonlocal advantage of quantum coherence (NAQC)\cite{Radwan2020} can  be
attained by the conditional state of subsystem $B$. The
  one norm coherence   $C_{l_{1}}^{na1}$  is defined as,
\begin{eqnarray}
\begin{aligned}
C_{l_{1}}^{na1}&=\sum_{i\neq j}|\langle i|\rho_{ij}|j \rangle|.
\end{aligned}
\end{eqnarray}
Thus, the criterion for achieving a NAQC on
qubit $B$ can be derived via the  possible probabilistic averaging methods \cite{cohe19}
\begin{eqnarray}\label{cooh2}
\tilde{C}_{\alpha}^{na}=\frac{1}{2}\sum_{\substack{i,j,a\\i\neq j}}p_{a|\prod_{i}^{a}}C_{\alpha}^{\sigma_{j}}(\rho_{B|\prod_{i}^{a}})>C_{\alpha}^{m},
\end{eqnarray}
where $ C_{\alpha}^{\sigma_{j}}(.) $($ \alpha=l_{1} $ ) represents the quantum coherence with respect to the reference basis spanned by the eigenstates of $ \sigma_{j} $, and the two critical values are given by $ C_{l_{1}}^{m} =\sqrt{6}$. The measured state for the two-qubit state may be described as,
\begin{equation}
\rho_{B|\prod_{i}^{a}}=\frac{Tr_{A}[(\prod_{i}^{a}\otimes I_{2})\rho_{AB}] }{p_{a|\prod_{i}^{a}}}.
\end{equation}
 Based on the criterion of equation (\ref{cooh2}), Hu et. al,\cite{cohe20} proposed to characterize quantitatively the degree of the non-local coherent advantage   $\mathcal{N}_{La}$
of   bipartite state  as,
\begin{equation}
\mathcal{N}_{La}(\rho_{AB})=\max\{0,\frac{\tilde{C}_{\alpha}^{na}(\rho_{AB})-C_{\alpha}^{m}}{\tilde{C}_{\alpha,\max}^{na}-C_{\alpha}^{m}}\},
\end{equation}
where $\tilde{C}_{\alpha,\max}^{na}=\max_{\rho_{AB}\in D(C^{d\times d})}\tilde{C}_{\alpha}^{na}(\rho_{AB})$ and  for the
two-qubit states we have  $ \tilde{C}_{\alpha,\max}^{na}=3 $.

\end{itemize}

\section{Numerical results}\label{Dip4}
\subsection{ Entangled two nodes}
In this subsection, we investigate the behavior of negativity, $\mathcal{N}_{eg}$ and the local  advantage coherent information, $\mathcal{N}_{La}$  of a channel  between  two different nodes. These nodes may be entangled from the beginning as $\rho_{12}$ and $\rho_{34}$, or direct/indirect  interacted nodes via Dipolar interaction as $\rho_{23}$ and $\rho_{14}$, respectively.

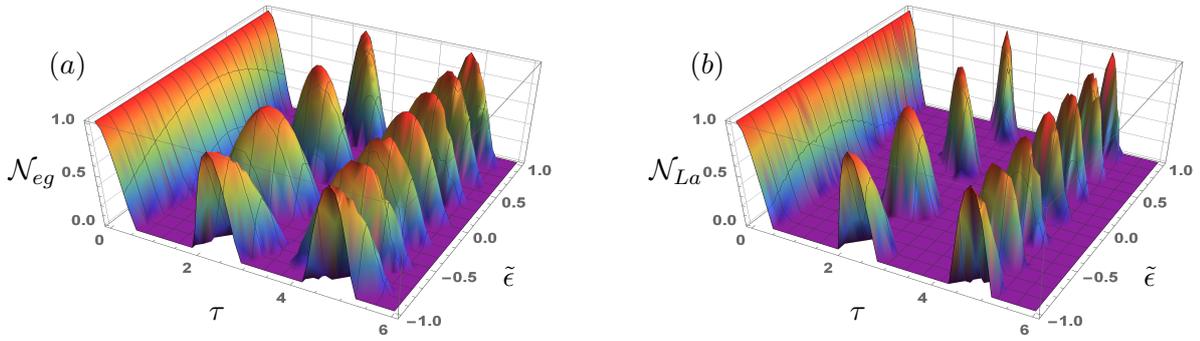
\begin{figure}[!h]
	\begin{center}
		\includegraphics[width=0.4\textwidth, height=125px]{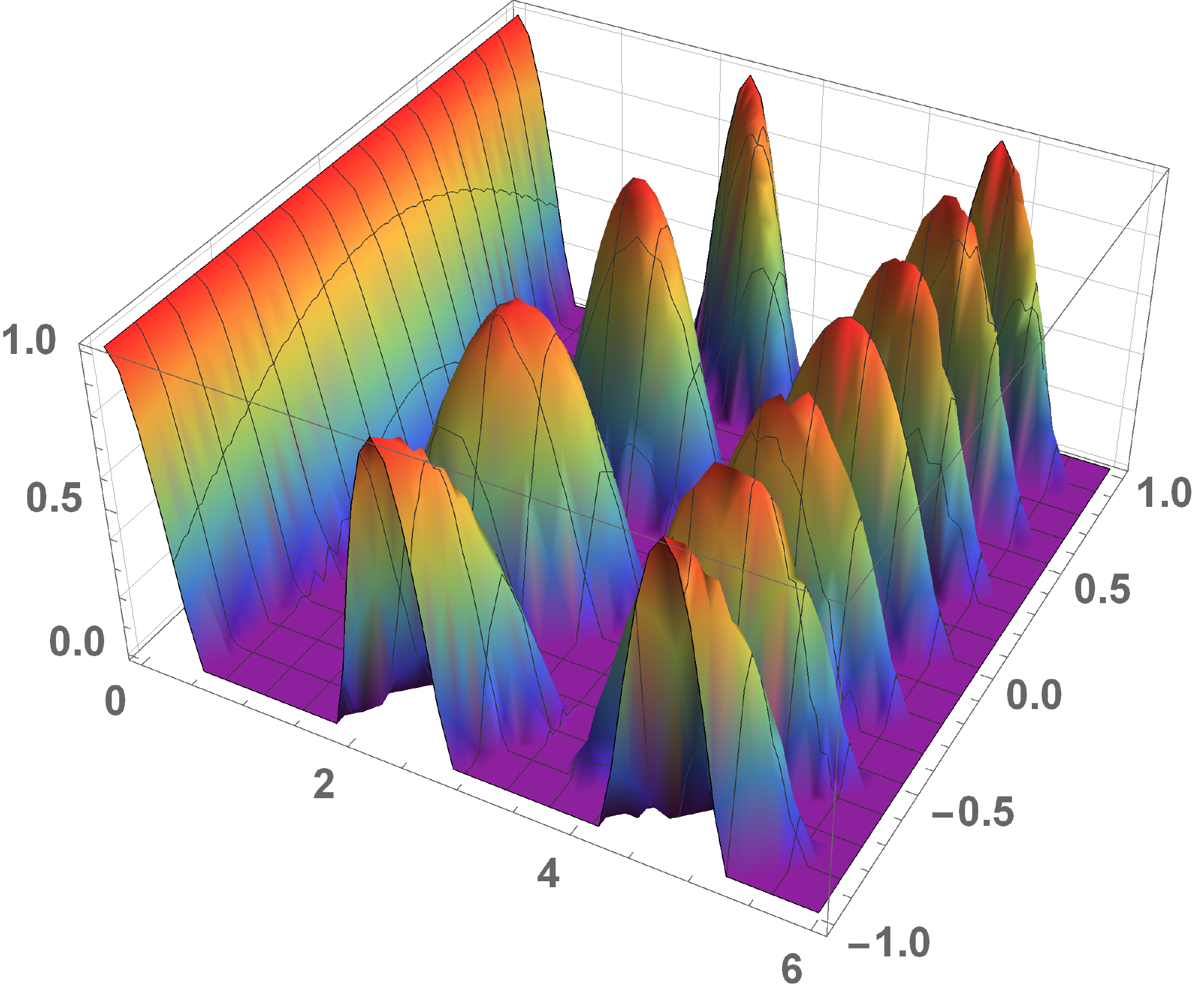}
		\put(-190,100){($ a $)}\put(-205,60){$\mathcal{N}_{eg}$}
		\put(-130,7){$\tau$}\put(-20,20){$\tilde{\epsilon}$}~~~~~\quad\quad\quad
\includegraphics[width=0.4\textwidth, height=125px]{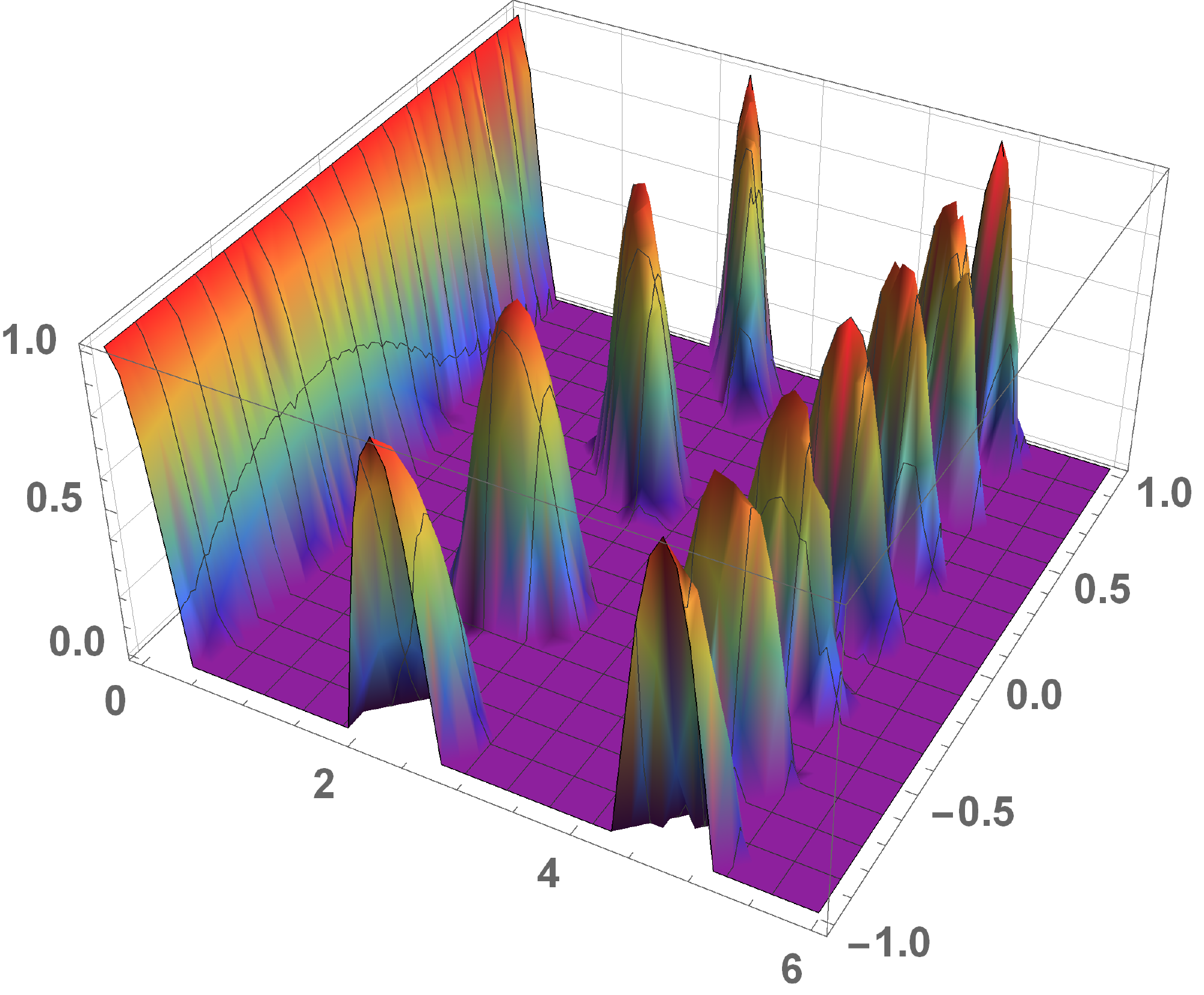}
		\put(-190,100){($ b $)}\put(-205,60){$\mathcal{N}_{La}$}
		\put(-130,7){$\tau$}\put(-20,20){$\tilde{\epsilon}$}~~~\quad\quad
	\end{center}
	\caption{\label{dipolar1}The behavior of the negativity, $\mathcal{N}_{eg}$ of $ \rho_{12} $ of the four qubits network as a function of $ \tau $ and $\tilde{\epsilon}$, where the network is initially of the type $MM$ network.}
\end{figure}
Fig.(\ref{dipolar1}) shows the  behavior of the negativity and the non-local coherent  advantage of  the quantum channel $\rho_{12}(t)$, where the initial network is prepared by using two singlet states. At $t=0$, both $\mathcal{N}_{eg}$ and $\mathcal{N}_{La}$ are maximum, where $\rho_{12}(0)$ is a maximum entangled state. However, as soon as the interaction is switched on, both  quantifiers decreases, suddenly to death completely. At further time, both of $\mathcal{N}_{eg}$ and $\mathcal{N}_{La}$ rebirth again to reach their maximum bounds periodically. From a geometrical perspective view of the these quantifiers, they behave  as peaks, where the area  of the base that predicted for the negativity is much larger than that displayed for the non-local coherent  advantage. This means that, the non-local coherent  advantage decreases faster than the negativity. Moreover, the number of peaks increases as the interaction time increases. The negative  values of $\bar\epsilon$, namely $\Delta<0$ means that, the Dipolar interaction is polarized in the $x-y$ plane, while the positive values of $\bar\epsilon$, i.e., $\Delta>0$ indicates the  Dipolar interaction is switched in the $z-$ axis. To display the effect of the negative and positive values of the coupling $\bar\epsilon$, we discuss the behavior of the $\mathcal{N}_{eg}$ and  $\mathcal{N}_{La}$ at some different values of $\bar\epsilon$.

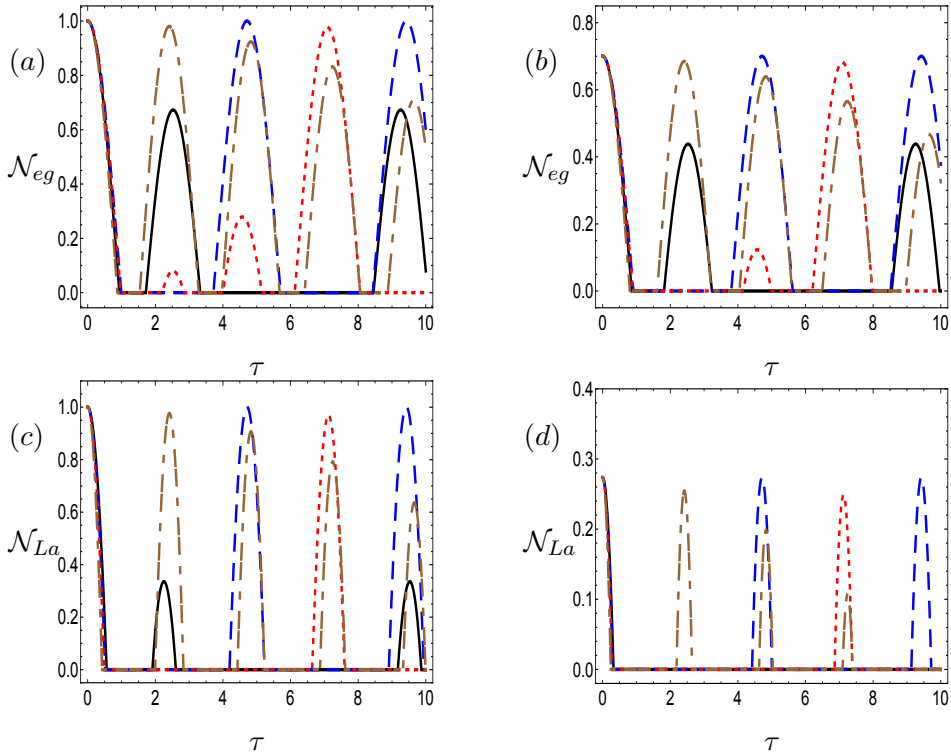
\begin{figure}[!h]
	\begin{center}
		\includegraphics[width=0.3\textwidth, height=125px]{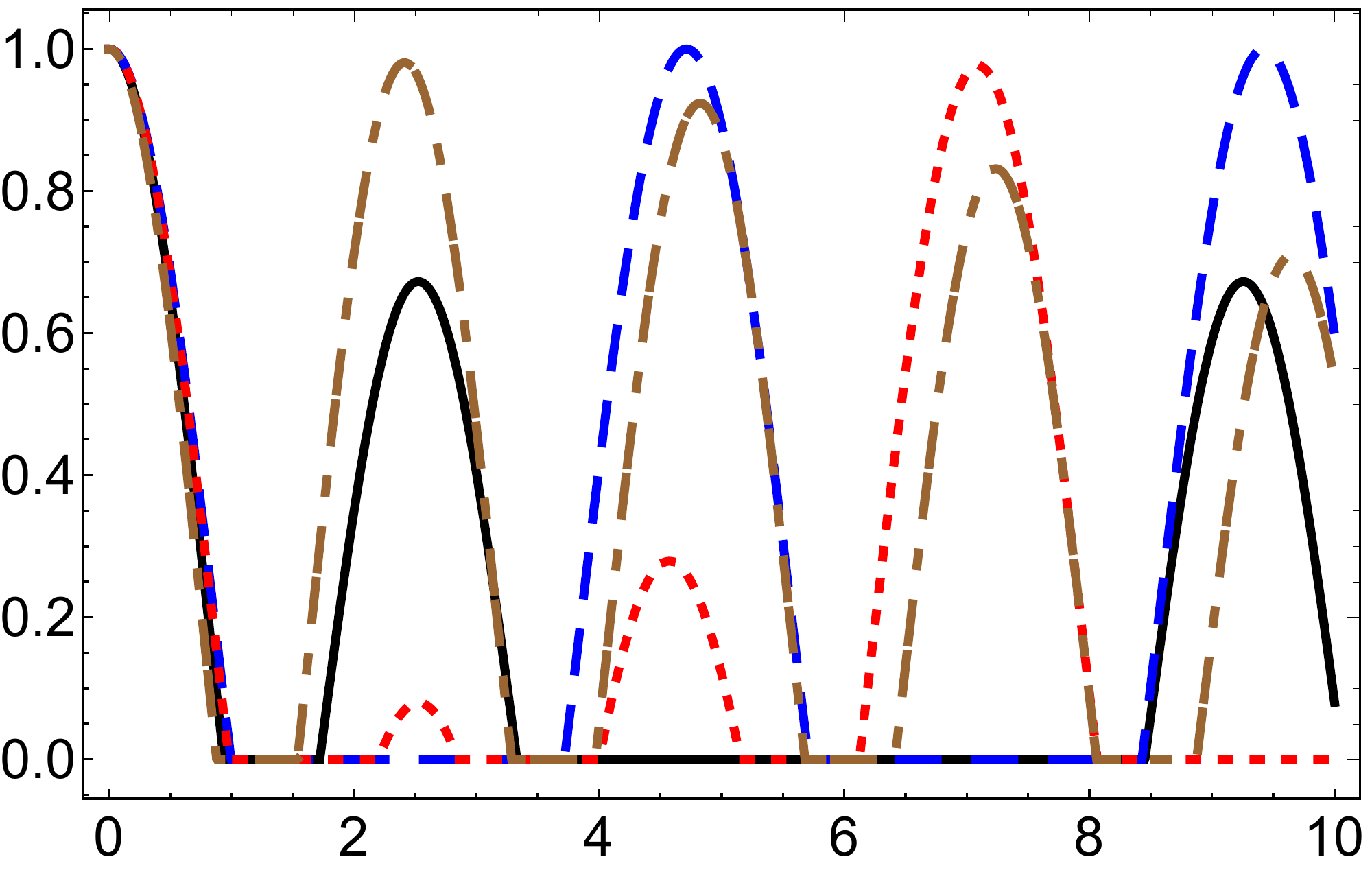}
		\put(-160,100){($ a $)}\put(-160,60){$\mathcal{N}_{eg}$}
		\put(-70,-15){$\tau$}~~~~~\quad\quad\quad
		\includegraphics[width=0.3\textwidth, height=125px]{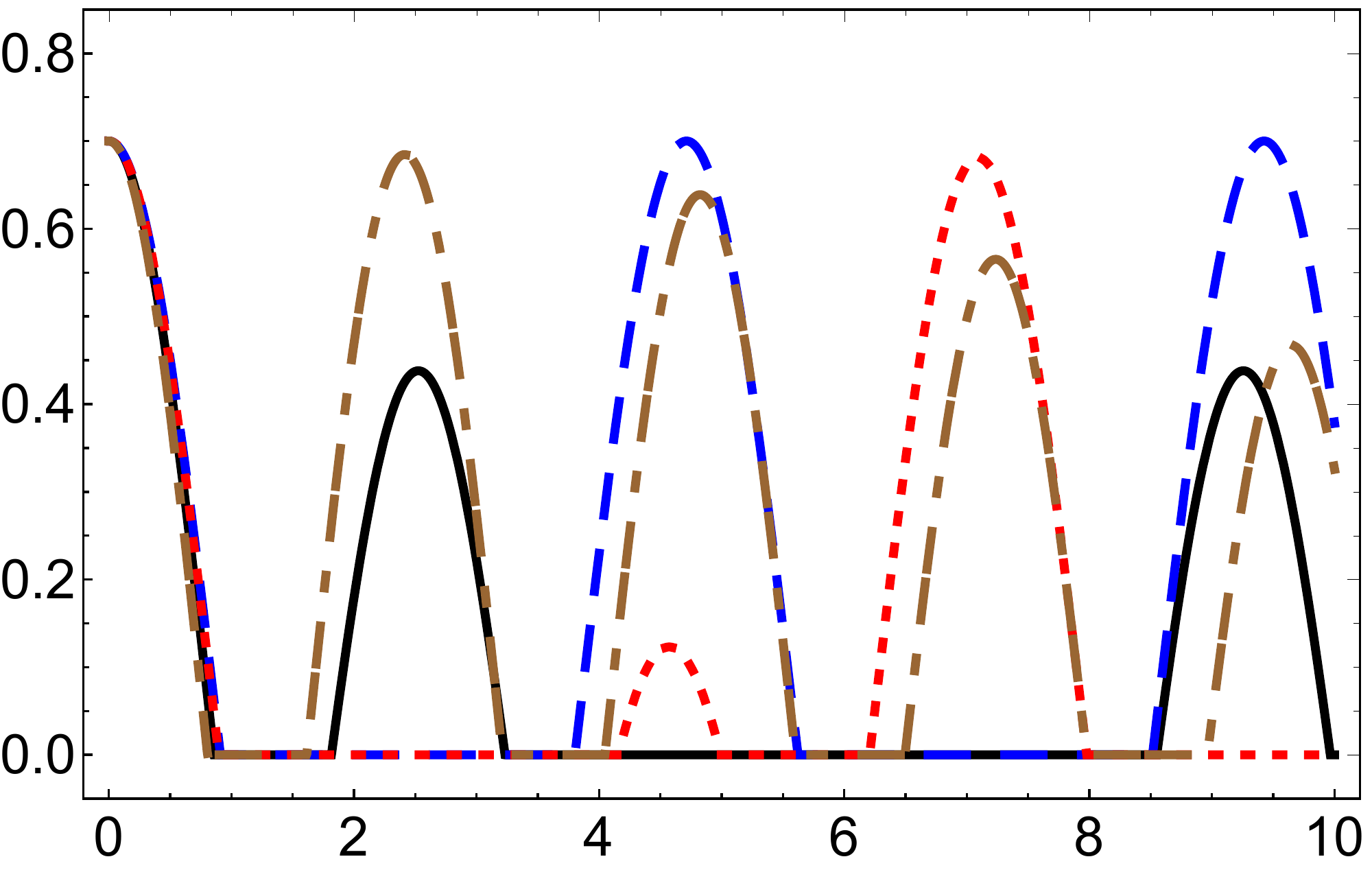}
		\put(-160,100){($ b $)}\put(-160,60){$\mathcal{N}_{eg}$}
		\put(-70,-15){$\tau$}~~~\quad\quad\\
\includegraphics[width=0.3\textwidth, height=125px]{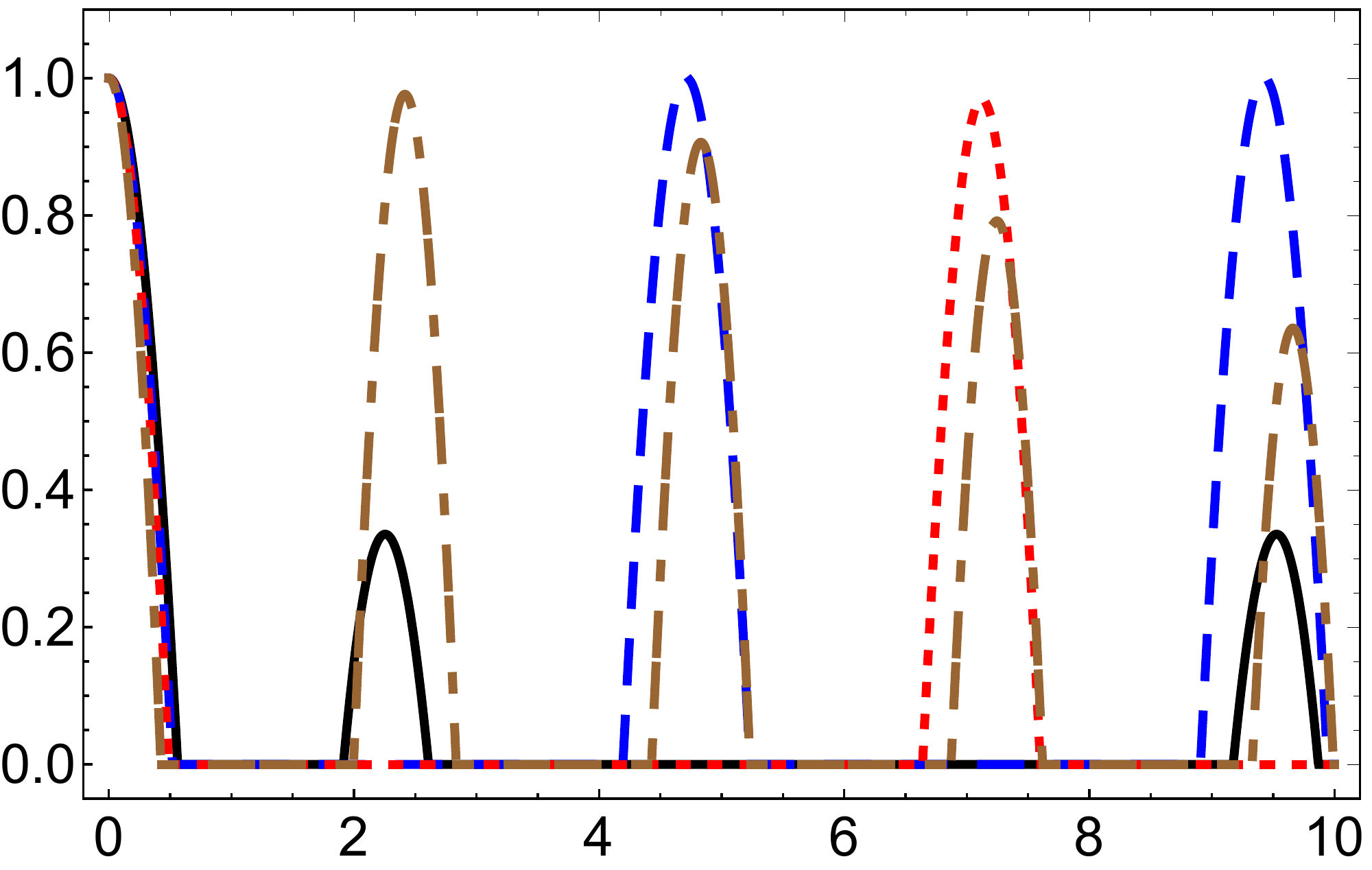}
		\put(-160,100){($ c $)}\put(-160,60){$\mathcal{N}_{La}$}
		\put(-70,-15){$\tau$}~~~~~\quad\quad\quad
		\includegraphics[width=0.3\textwidth, height=125px]{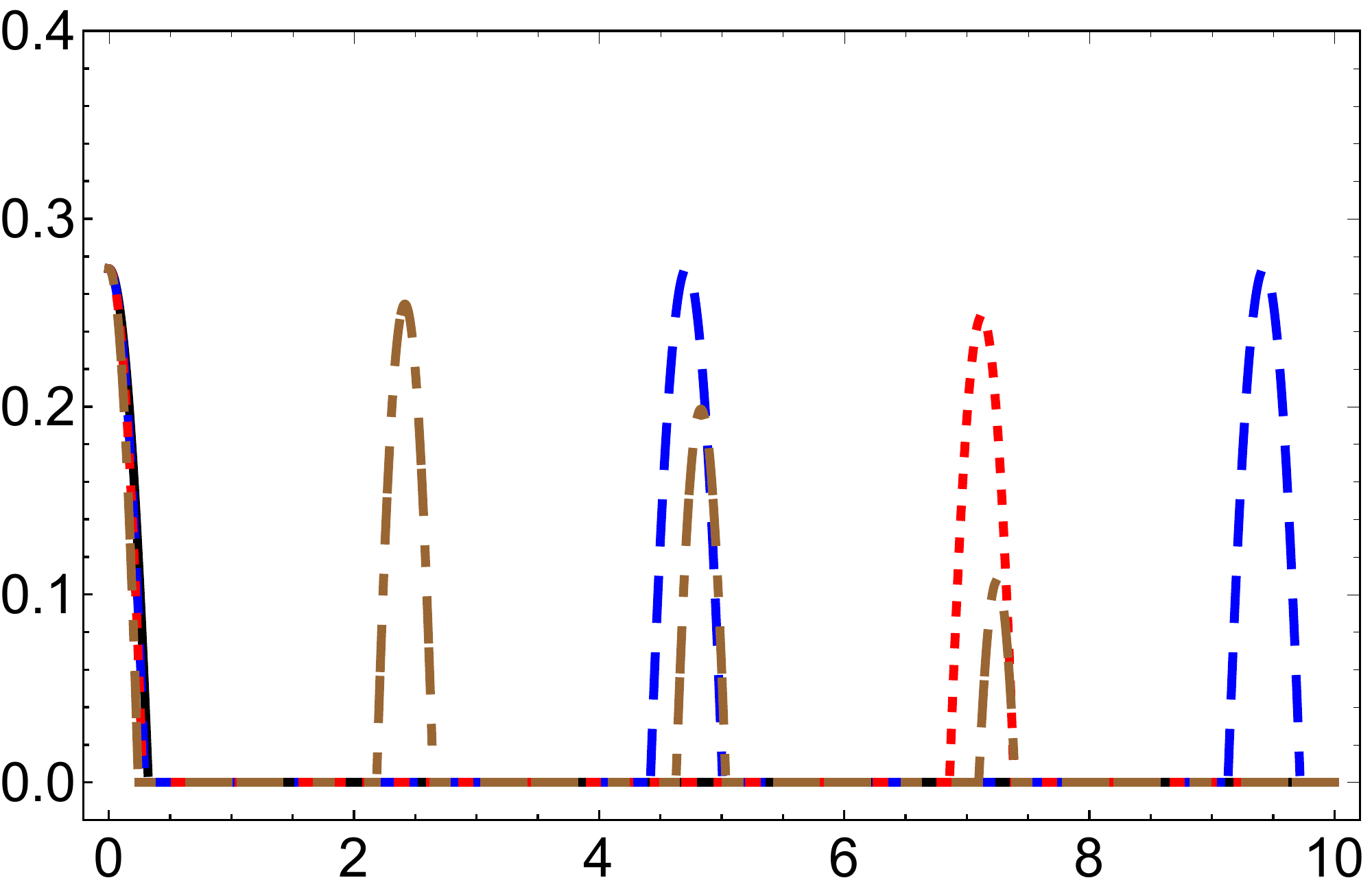}
		\put(-160,100){($ d $)}\put(-160,60){$\mathcal{N}_{La}$}
		\put(-70,-15){$\tau$}~~~\quad\quad
	\end{center}
\caption{\label{dipolar2}The behavior of the negativity  $\mathcal{N}_{eg}$ is represented by (a,b) and the non-local  coherent advantage, $\mathcal{N}_{La}$ is shown  in (c,d)of $ \rho_{12} $. The solid, dash, the dot and the dash-dot lines  are evaluated  at $\tilde{\epsilon}=-0.2, 0, 0.1, 0.3$, respectively. It is assumed that,  the network is initially of type $MM$ network   while it is of $WW$ network in $(a,c)$. }
\end{figure}

In Fig.(\ref{dipolar2}), we display the sensitive of both quantifiers  to the positive/negative   values of the coupling $\bar\epsilon$, where it is assumed that, the initial network is constructed either from maximum or partial entangled nodes.  The general behavior of both quantifiers displays the phenomena of the sudden death/birth of the negativity and the non-local coherent advantage. The death time that  is shown for the negativity is smaller than  that displayed for the non-local  coherent advantage. Moreover,  the maximum values of  $\mathcal{N}_{eg}$  and $\mathcal{N}_{La}$ that predict in Figs.(a,c) are much  larger  than those displayed in Figs.(b,d), where the initial network is  of $MM$ and $WW$ types, respectively.

\subsection{Indirect entangled nodes}
As  it is  mentioned above, due to the interaction there are some    entangled  two nodes channels  are generated. In this subsection, we quantify the amount of entanglement, as well as, the non-local coherent advantage that contained in the  quantum channel between the first  and the forth nodes, namely $\rho_{14}$.
\begin{figure}[!h]
	\begin{center}
		\includegraphics[width=0.4\textwidth, height=125px]{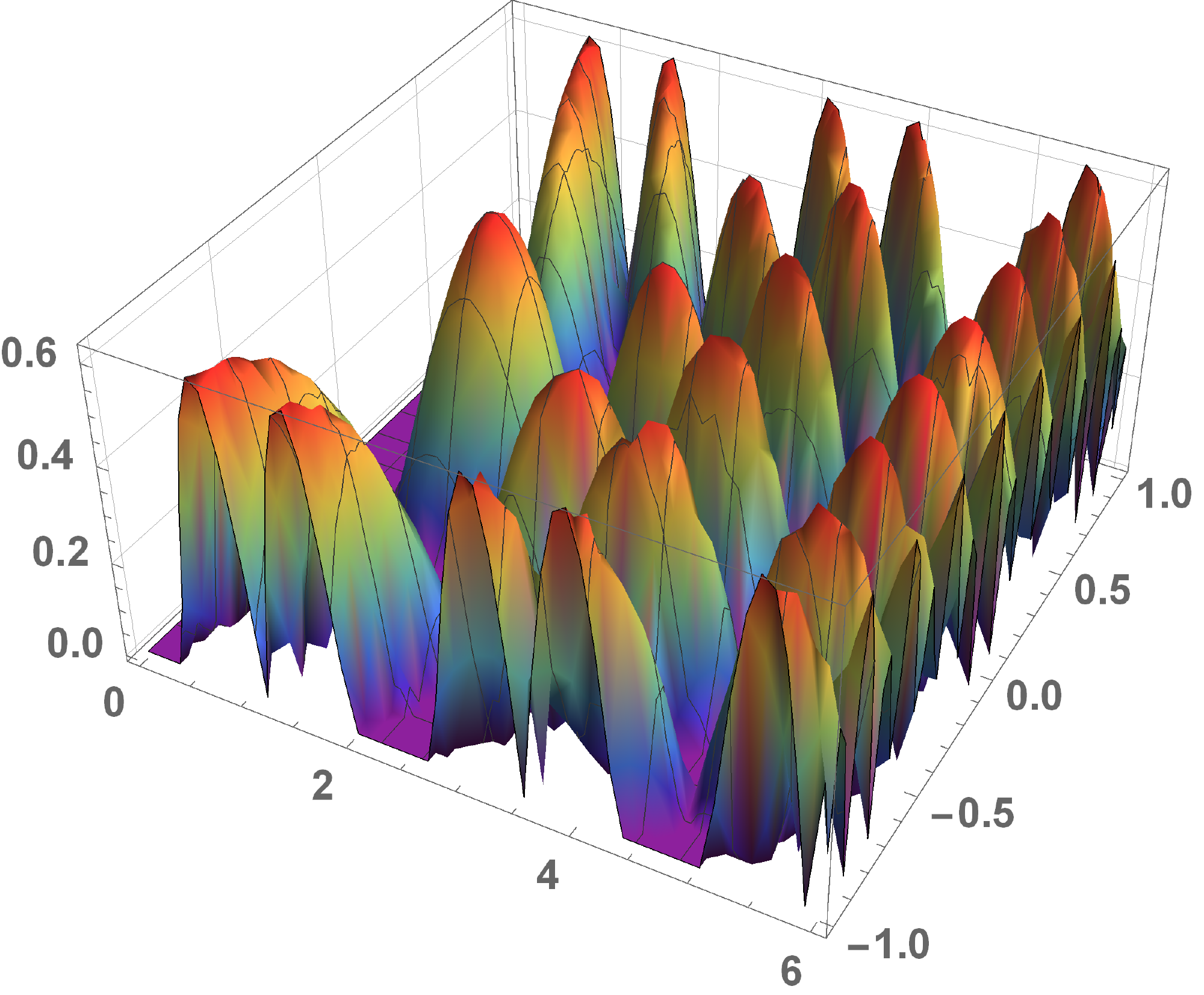}
		\put(-190,100){($ a $)}\put(-205,60){$\mathcal{N}_{eg}$}
		\put(-130,7){$\tau$}\put(-20,20){$\tilde{\epsilon}$}~~~\quad\quad
\includegraphics[width=0.4\textwidth, height=125px]{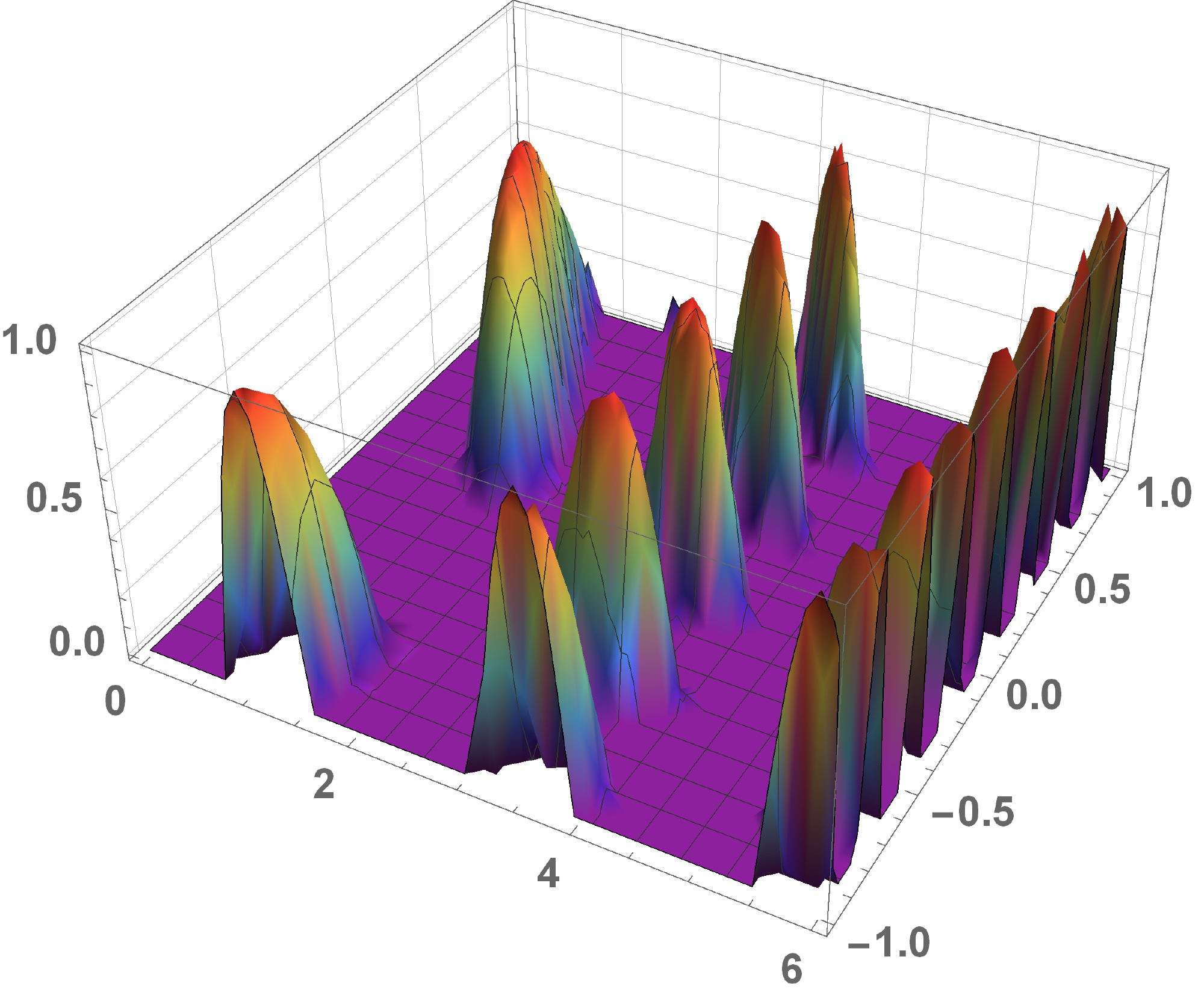}
		\put(-190,100){($ a $)}\put(-205,60){$\mathcal{N}_{La}$}
		\put(-130,7){$\tau$}\put(-20,20){$\tilde{\epsilon}$}~~~\quad\quad
	\end{center}
	\caption{\label{dipolar3} The same as Fig.(\ref{dipolar1}), but for the two nodes $\rho_{14}.$}
\end{figure}
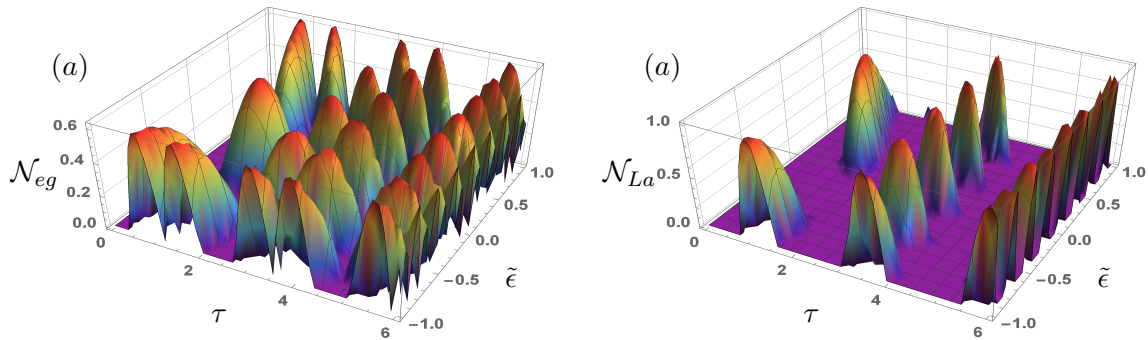
Fig.(\ref{dipolar3}), describes the behavior of  both quantifiers  for the  generated entangled  two  nodes channels $\rho_{14}$, where we assume  that, the initial network is  of $MM$ type.  As it is displayed from Fig.(\ref{dipolar3}a), the behavior of the negativity $\mathcal{N}_{eg}$ is similar to that displayed in Fig.(\ref{dipolar1}a). However, the number of picks are  larger than those displayed  for the channel nodes $\rho_{12}$. Moreover, the maximum bounds of entanglement  that displayed for $\rho_{14}$ are smaller than those displayed for the  channel $\rho_{12}$. Fig.(\ref{dipolar3}b), describes the behavior of the non-local coherent advantage for the channel $\rho_{14}$. The behavior   of $\mathcal{N}_{La}$  is similar to that shown in Fig.(\ref{dipolar1}a). However, the number of peaks is larger than those  displayed  for the channel $\rho_{12}$. Moreover, the behavior of both  quantifiers  shows  that, the peaks  of  $\rho_{14}$  appear faster than those displayed for the channel $\rho_{12}$.

From Fig.(\ref{dipolar1}) and (\ref{dipolar3}), one may conclude that, the entanglement   that generated indirectly is more stable than that between the initial entangled nodes, where the vanishing time of both quantifiers that depicted for the indirect connected nodes   is much smaller than that  displayed for the initially entangled.

\begin{figure}[!h]
	\begin{center}
		\includegraphics[width=0.3\textwidth, height=125px]{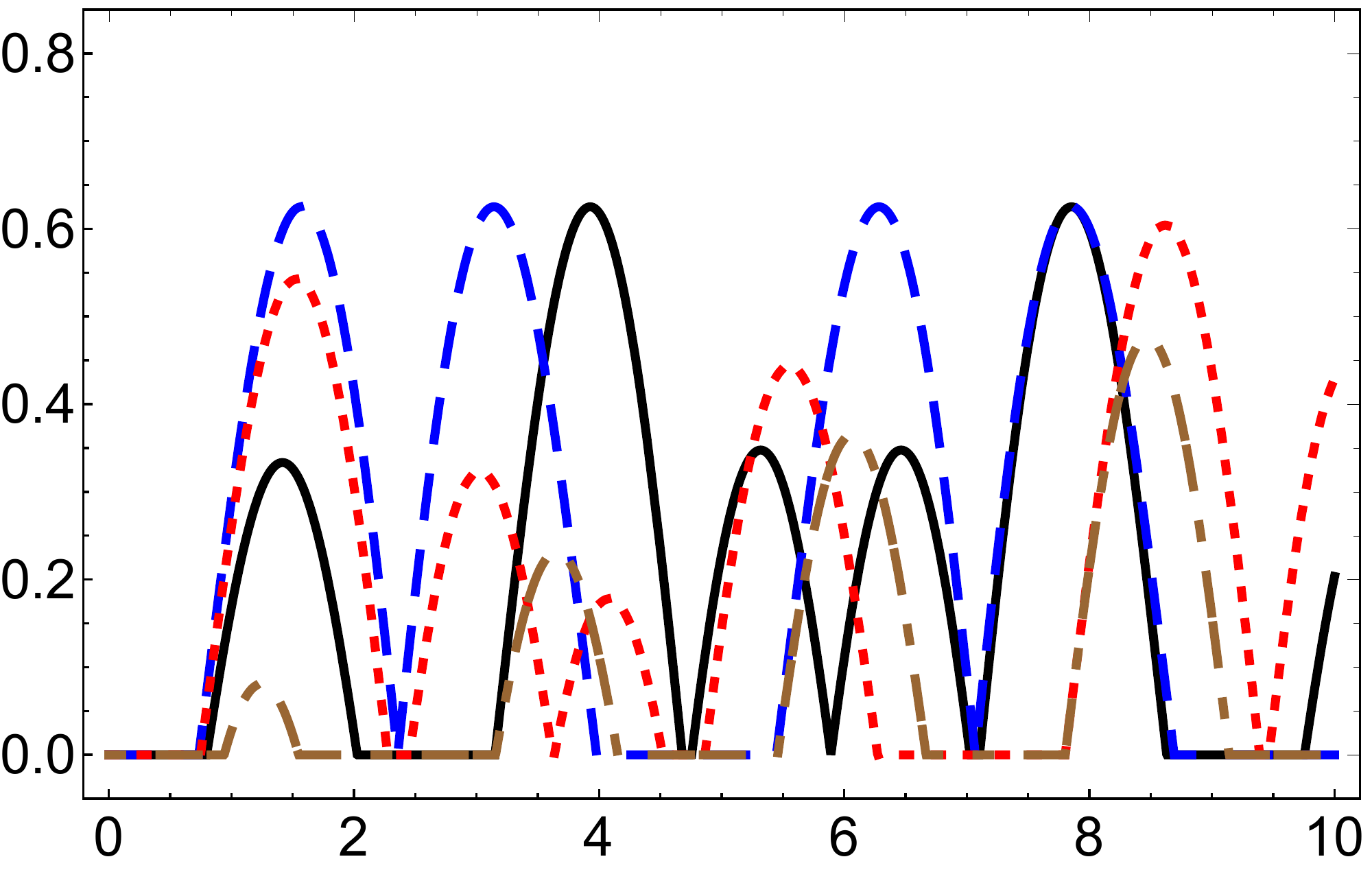}
		\put(-160,100){($ a $)}\put(-160,60){$\mathcal{N}_{eg}$}
		\put(-70,-15){$\tau$}~~~\quad\quad
		\includegraphics[width=0.3\textwidth, height=125px]{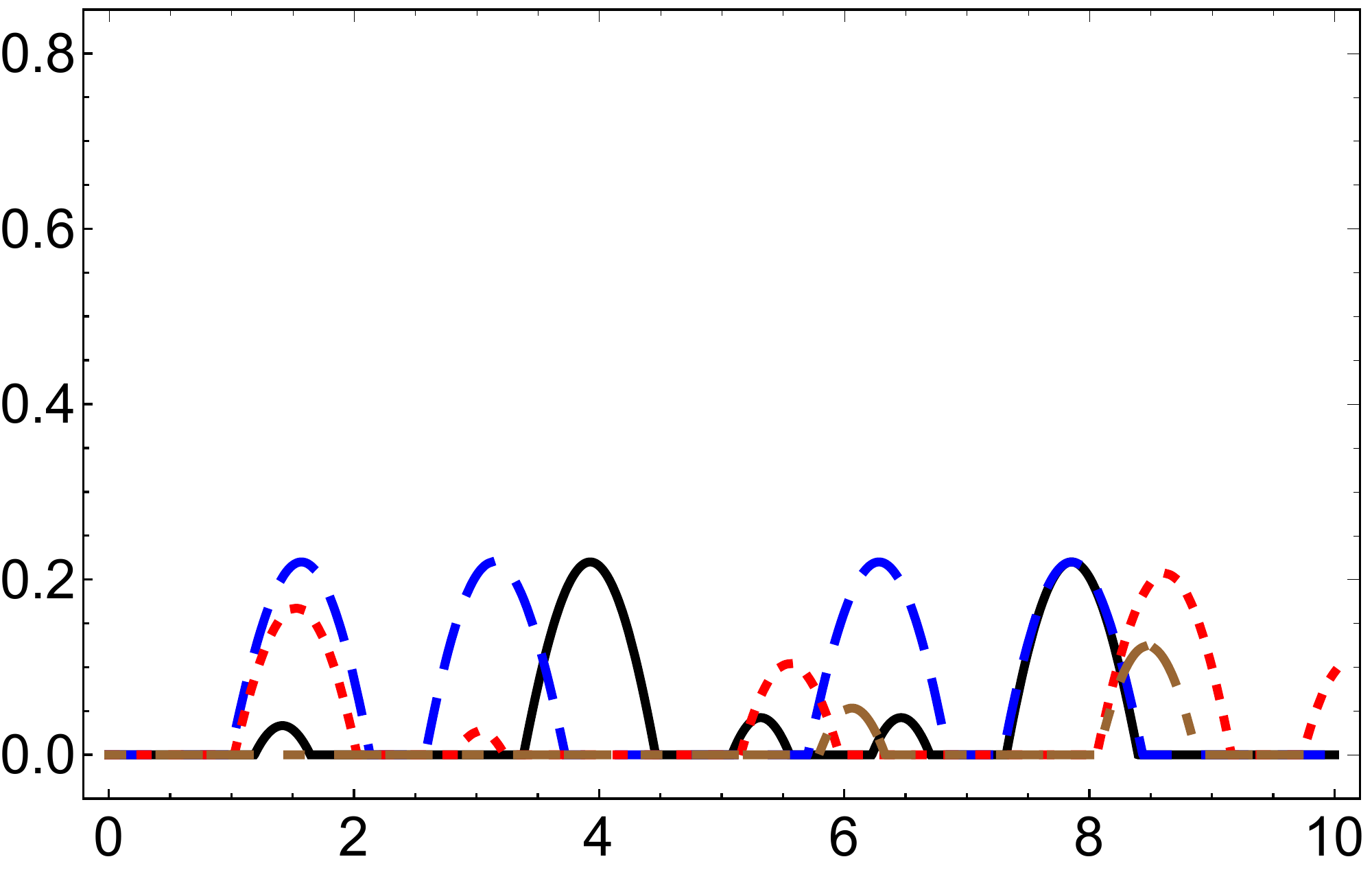}
		\put(-160,100){($ b $)}\put(-160,60){$\mathcal{N}_{eg}$}
		\put(-70,-15){$\tau$}~~~\quad\quad
		\includegraphics[width=0.3\textwidth, height=125px]{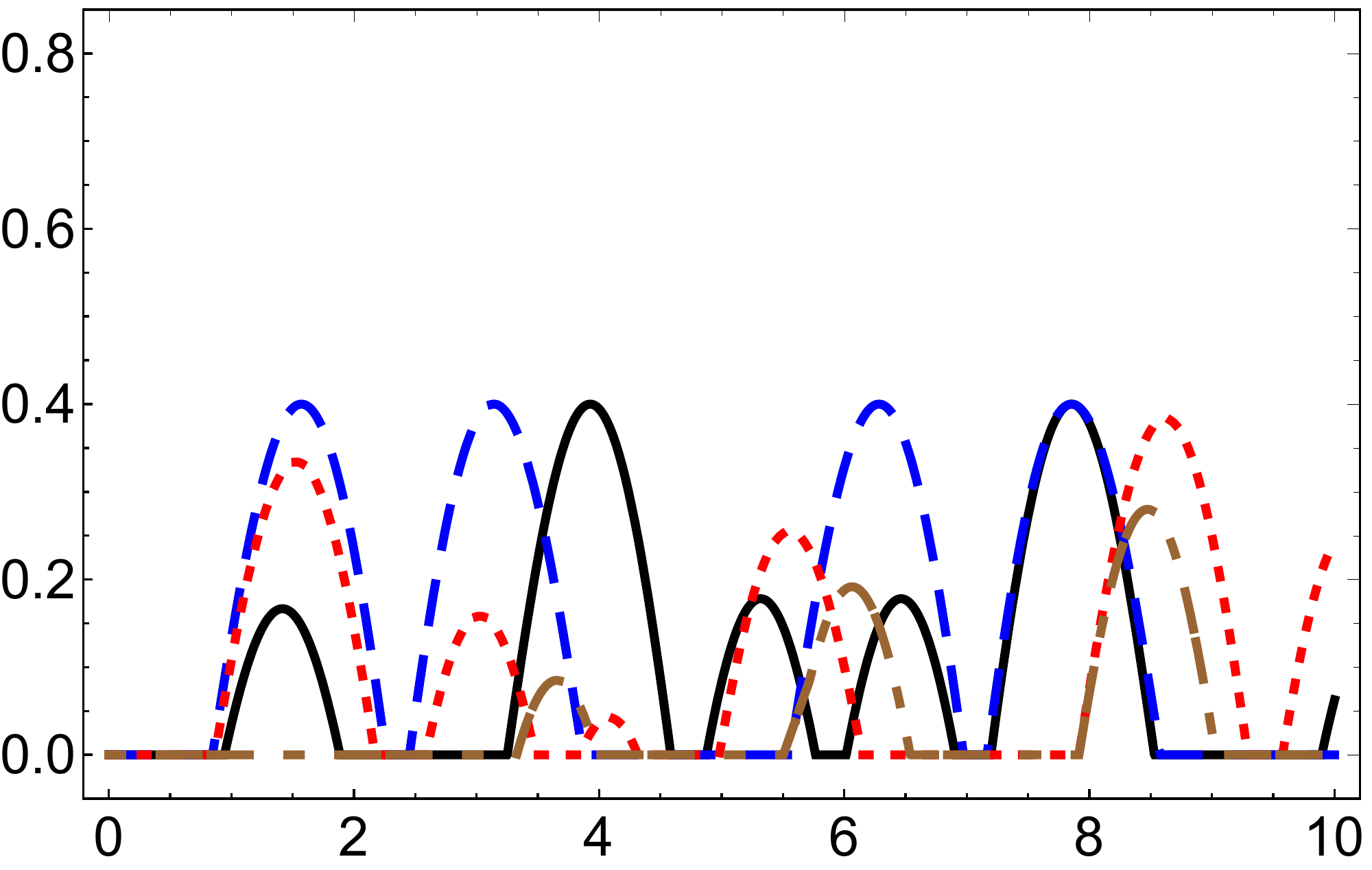}
		\put(-160,100){($ c $)}\put(-160,60){$\mathcal{N}_{eg}$}
		\put(-70,-15){$\tau$}~~~\quad\quad
	\end{center}
	\caption{\label{dipolar4}The  negativity behavior of $ \rho_{14} $, where  different values of the coupling constant are considered. The solid, dash, the dot and the dash-dot curves represents the   $\mathcal{N}_{eg}$ at  $\tilde{\epsilon}=-0.2, 0, 0.1, 0.3$, respectively  (a) $MM$ network, (b) $ WW$ network, and  (c)$ MW$ network. }
\end{figure}

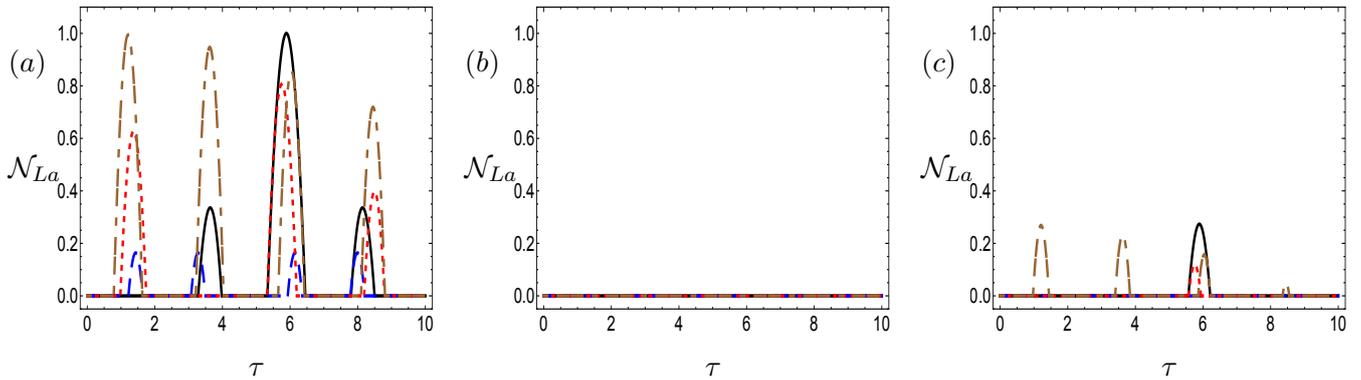
\begin{figure}[!h]
	\begin{center}
		\includegraphics[width=0.3\textwidth, height=125px]{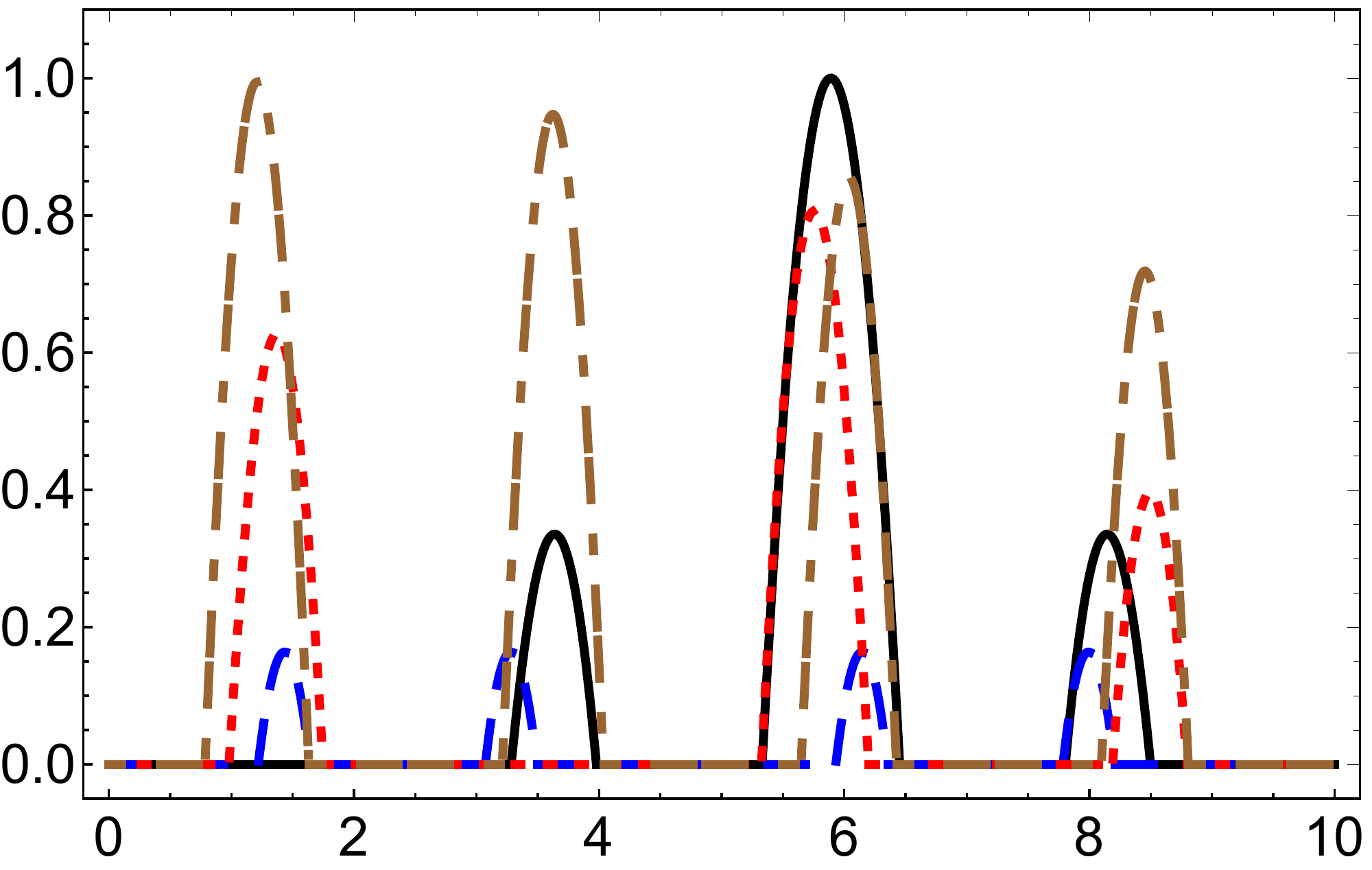}
		\put(-160,100){($ a $)}\put(-160,60){$\mathcal{N}_{La}$}
		\put(-70,-15){$\tau$}~~~\quad\quad
		\includegraphics[width=0.3\textwidth, height=125px]{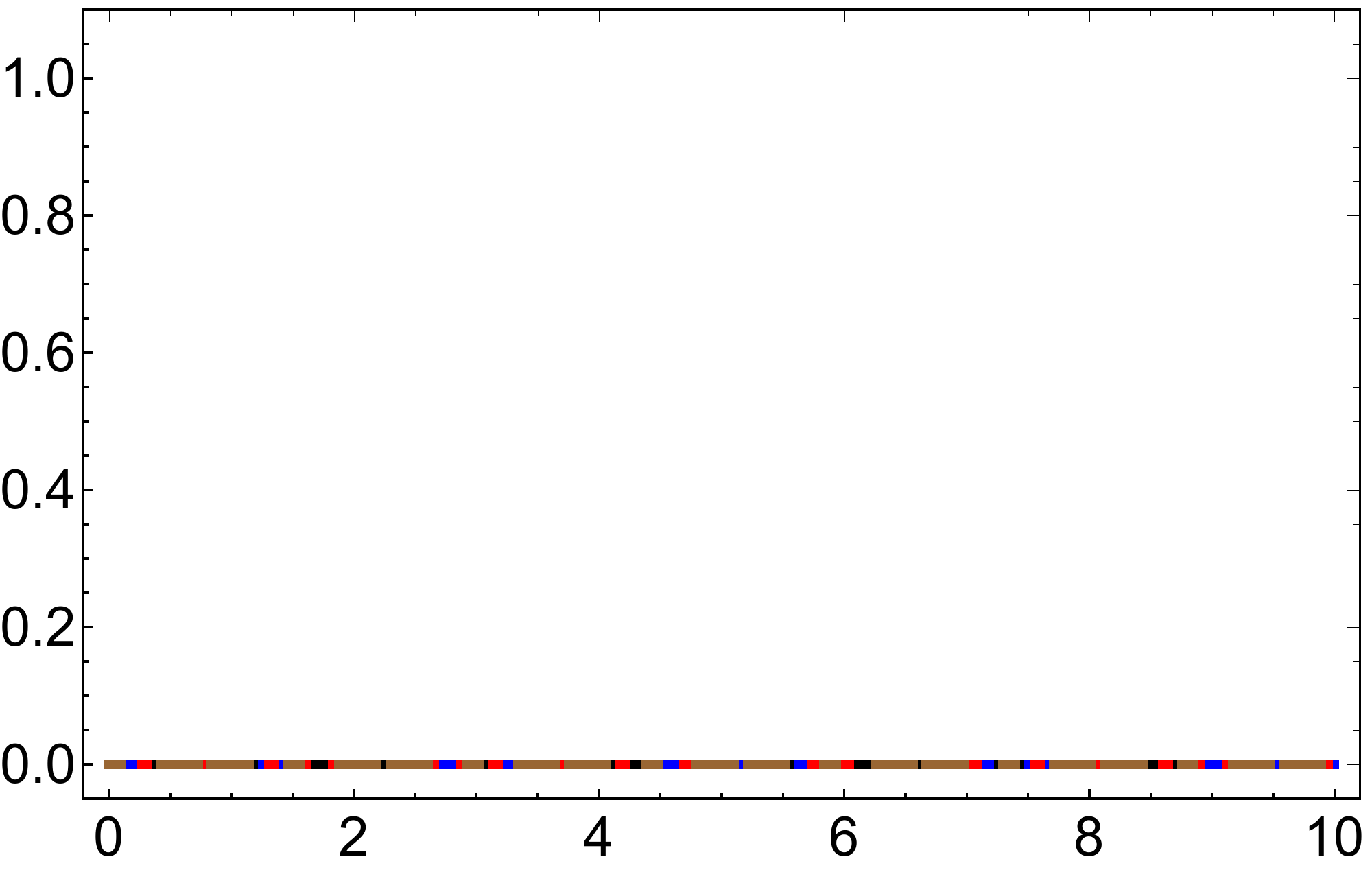}
		\put(-160,100){($ b $)}\put(-160,60){$\mathcal{N}_{La}$}
		\put(-70,-15){$\tau$}~~~\quad\quad
		\includegraphics[width=0.3\textwidth, height=125px]{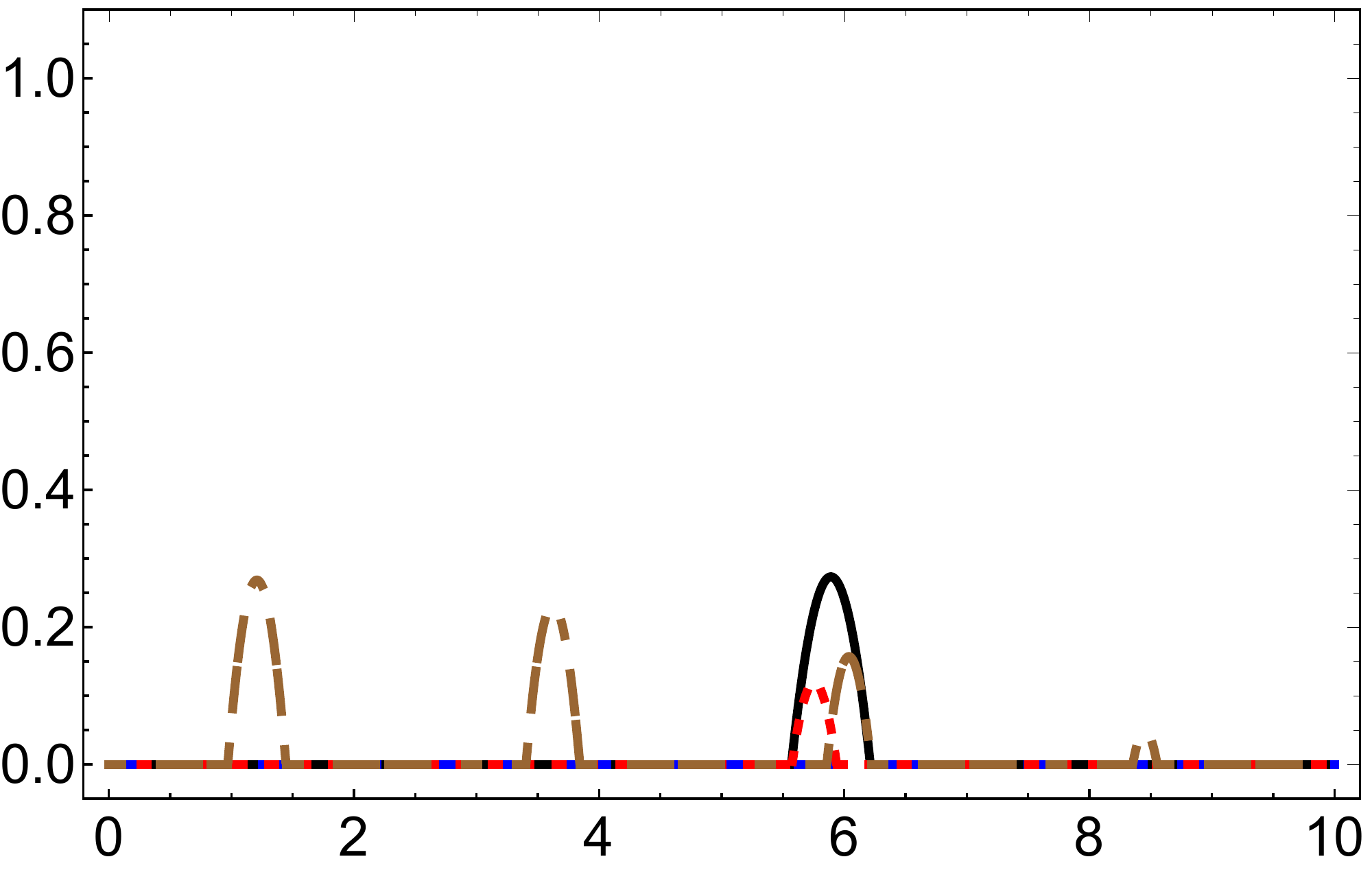}
		\put(-160,100){($ c $)}\put(-160,60){$\mathcal{N}_{La}$}
		\put(-70,-15){$\tau$}~~~\quad\quad
	\end{center}
	\caption{\label{dipolar5}The same as Fig.(\ref{Dip4}) but for the non-local coherent advantage, $\mathcal{N}_{La}$ of $ \rho_{14} $.}
\end{figure}

In Fig.(\ref{dipolar4}), we examine the effect considered  some values of the coupling strength $\tilde\epsilon$, and different settings of the initial networks on the negativity of the entangled nodes $\rho_{14}$. The behavior is similar to that shown in Fig.(\ref{dipolar2}). However,   at $t>0$, the entanglement reaches its maximum values faster and
it appears for a short time. The maximum bounds of $\mathcal{N}_{eg}$  that depicted at  the different values of $\tilde\epsilon$  are similar. It is clear that, at $\tilde\epsilon=0$, and $\Delta$ could be arbitrary, the sudden time   death   of the entanglement is very short and the entanglement oscillates only between $0$ and the maximum values $(\mathcal{N}_{eg}\simeq 0.6)$.  However, the smallest values of the entanglement are displayed  when the Dipolar interaction is switched on the $z-$ axis, namely $\Delta>0$.
The initial entangled network has   a clear effect on the behavior of the generated entangled nodes. It is clear that,  if the network is initially is of $MM$ type, the  amount of entanglement that generated between the second and the third nodes is  the largest one. However, the amount of entanglement on the channel $\rho_{23} $ is the smallest one, when the initial network is prepared by using partially entangled nodes $WW-$type. These results may be seen clearly, by comparing  Figs.(\ref{dipolar4}a)(\ref{dipolar4}b) and (\ref{dipolar4}c), where the network is initially  prepared by using maximum-maximum entangled  nodes (MM), Werner-Werner-entangled nodes (WW) or maximum-Werner entangled nodes (MW), respectively.

The behavior of the non-local coherent advantage  of the channel  between the first and the fourth node, $\rho_{14}$ is shown in Fig.(6). The behavior  shows that,  it fluctuates between its maximum and minimum values. The sudden changes of $\mathcal{N}_{La}$ (death/rebirth)  appear clearly, where the death time of this phenomena depends on the strength of the Dipolar interaction. Moreover,  the amount of  correlation that quantified  by $\mathcal{N}_{La}$ depends on the initial type of the network. It is clear that, there is no any correlations are predicted  during the same period of interaction time, when the network is initially prepared by using $WW$ entangled nodes. However, the maximum values of the non-local coherent advantage are displayed when  the initial network is conducted  by using maximum entangled nodes  ($MM$-type).

\subsection{Direct entangled nodes}
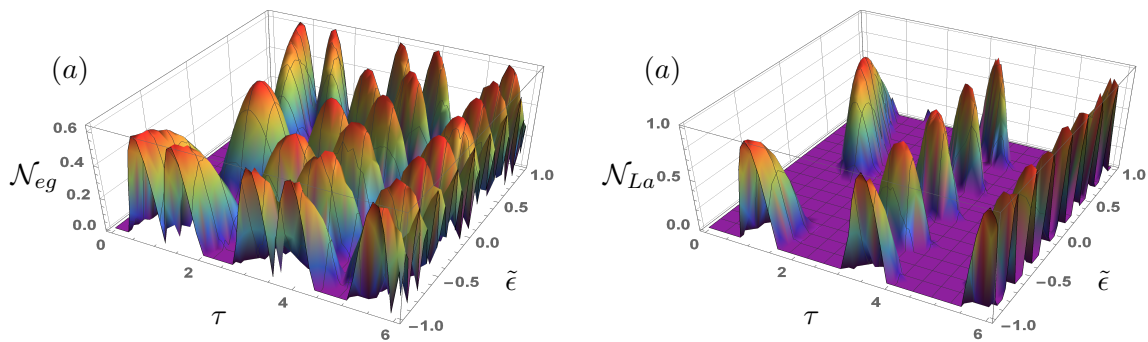
\begin{figure}[!h]
	\begin{center}
		\includegraphics[width=0.4\textwidth, height=125px]{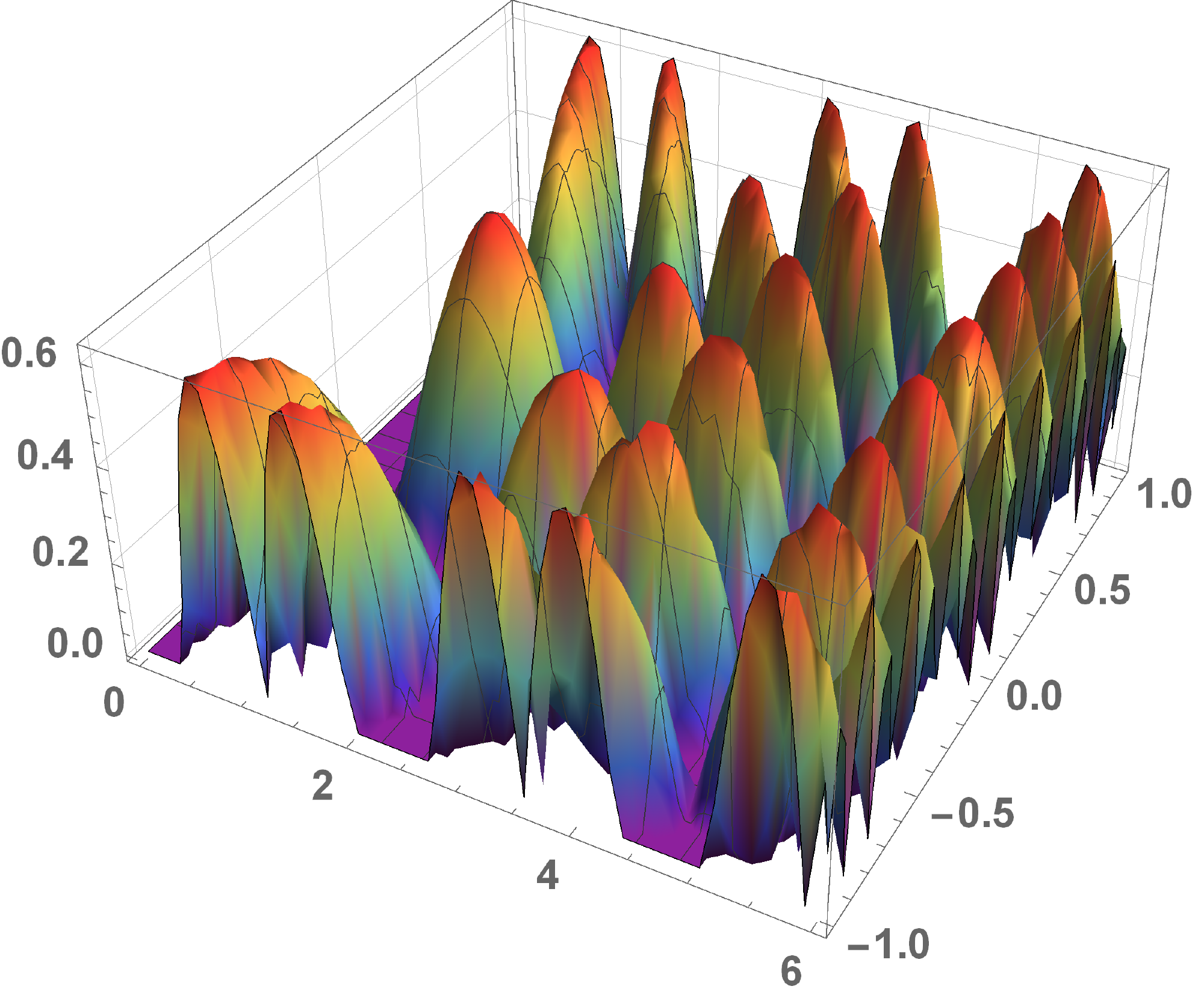}
		\put(-190,100){($ a $)}\put(-205,60){$\mathcal{N}_{eg}$}
		\put(-130,7){$\tau$}\put(-20,20){$\tilde{\epsilon}$}~~~\quad\quad
\includegraphics[width=0.4\textwidth, height=125px]{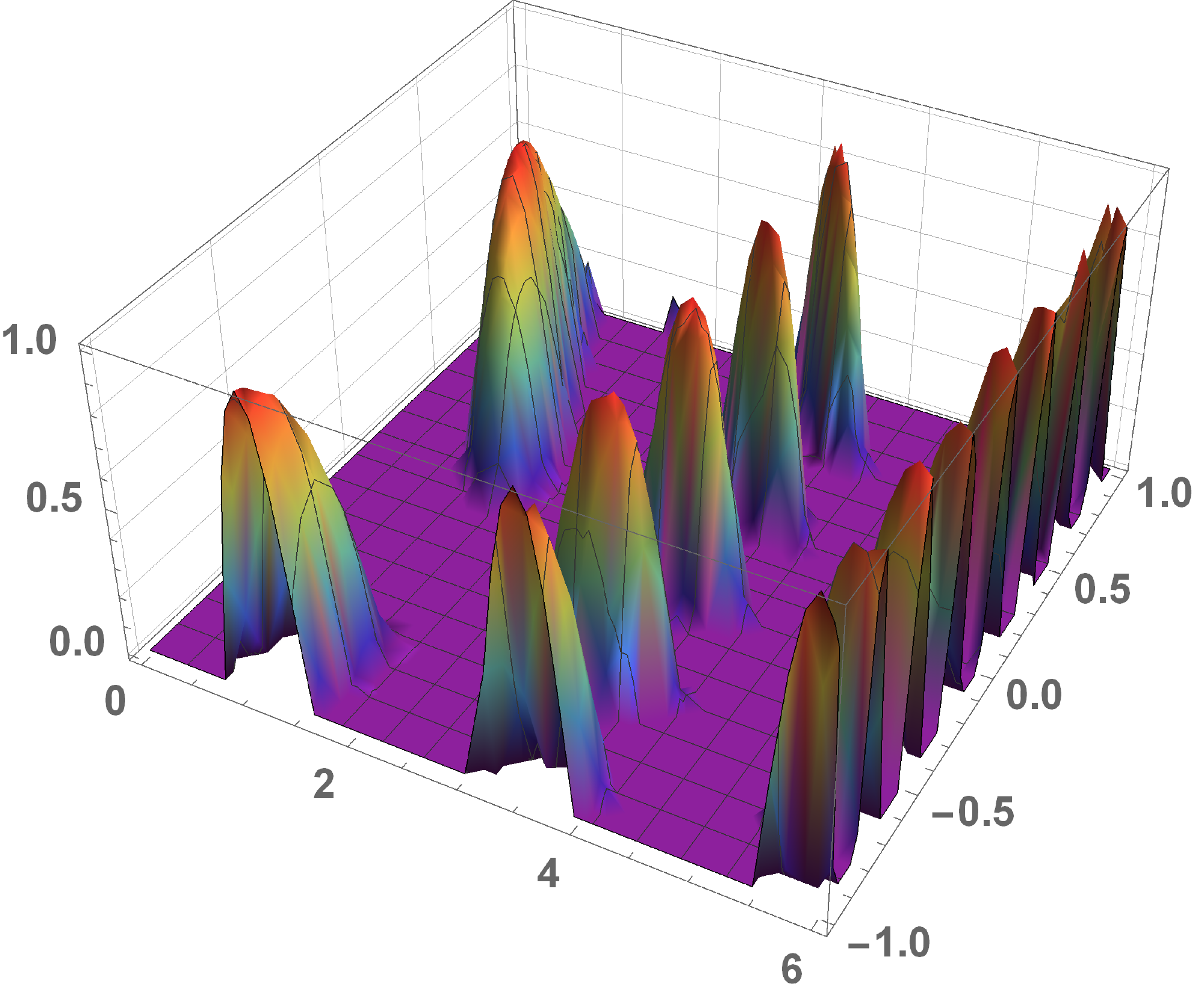}
		\put(-190,100){($ a $)}\put(-205,60){$\mathcal{N}_{La}$}
		\put(-130,7){$\tau$}\put(-20,20){$\tilde{\epsilon}$}~~~\quad\quad
	\end{center}
	\caption{\label{dipolar6}(a) The behavior of the negativity,and (b) the non-local coherent advantage  of the state $ \rho_{23} $ of the four qubits network as a function of $ \tau $ and $\tilde{\epsilon}$, where the initial network  is  $MM$- type.}
\end{figure}

The behavior of the entanglement and the non-local coherent advantage  that  are displayed  for  the entangled nodes state $\rho_{23}$  by using direct interaction is shown in Fig.(\ref{dipolar6}), where it is assumed that, the initial network is superimposed  from   maximum entangled nodes ($MM$-type). The behavior is similar to that displayed in Figs.(\ref{dipolar3}), where the sudden death phenomena of the non-local coherent advantage appears periodically during a  large period of interaction time.
Moreover, the sudden death phenomenon of the  non local coherent advantage decreases as the interaction time increases.

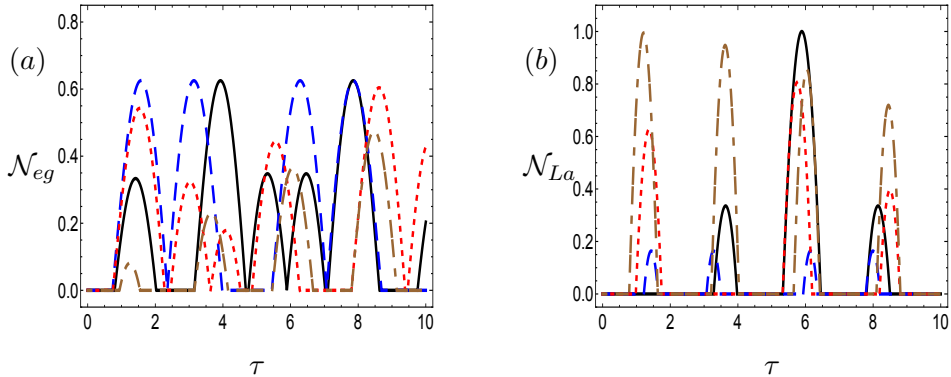
\begin{figure}[!h]
	\begin{center}
		\includegraphics[width=0.3\textwidth, height=125px]{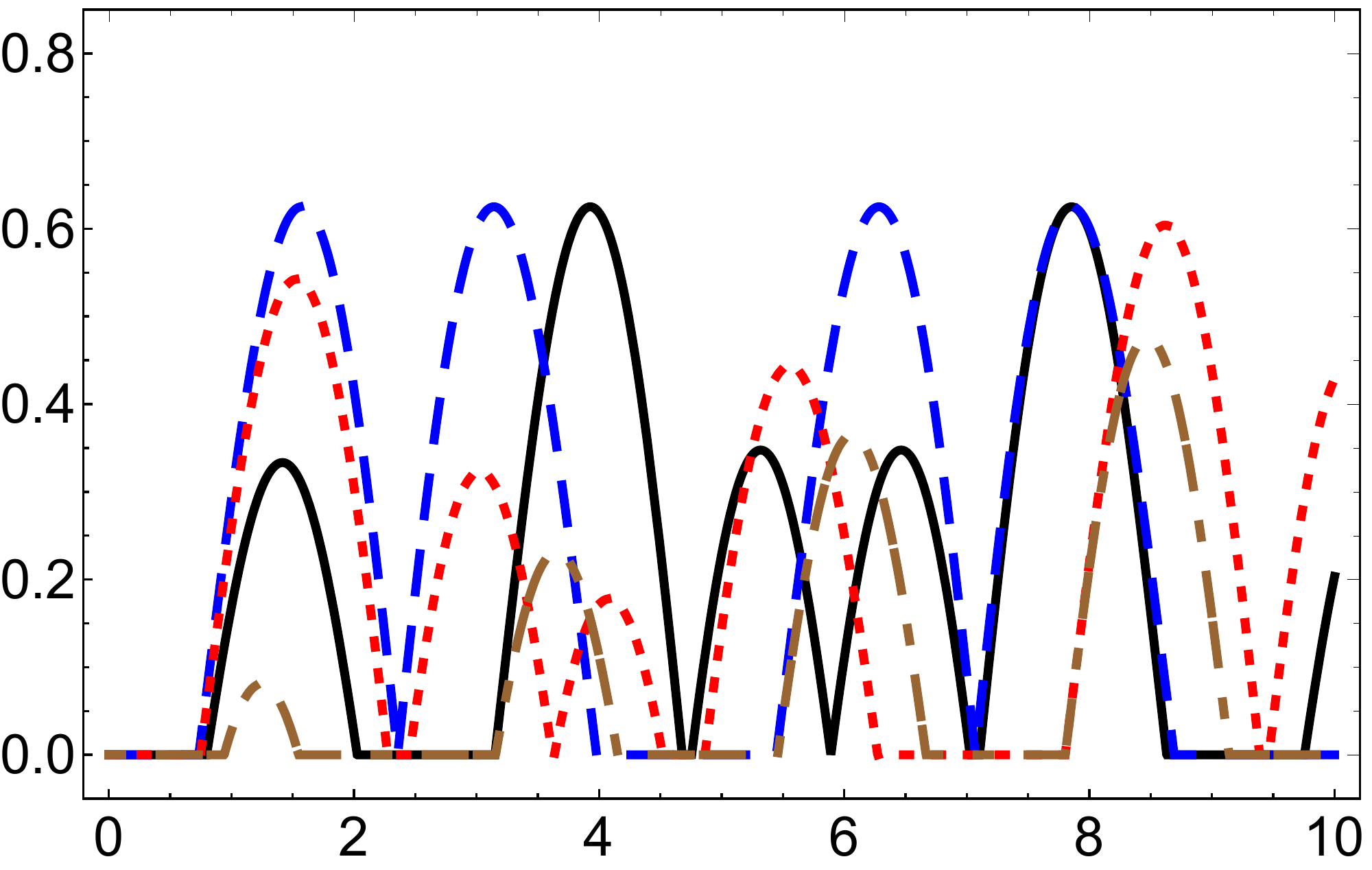}
		\put(-160,100){($ a $)}\put(-160,60){$\mathcal{N}_{eg}$}
		\put(-70,-15){$\tau$}~~~~~\quad\quad\quad
\includegraphics[width=0.3\textwidth, height=125px]{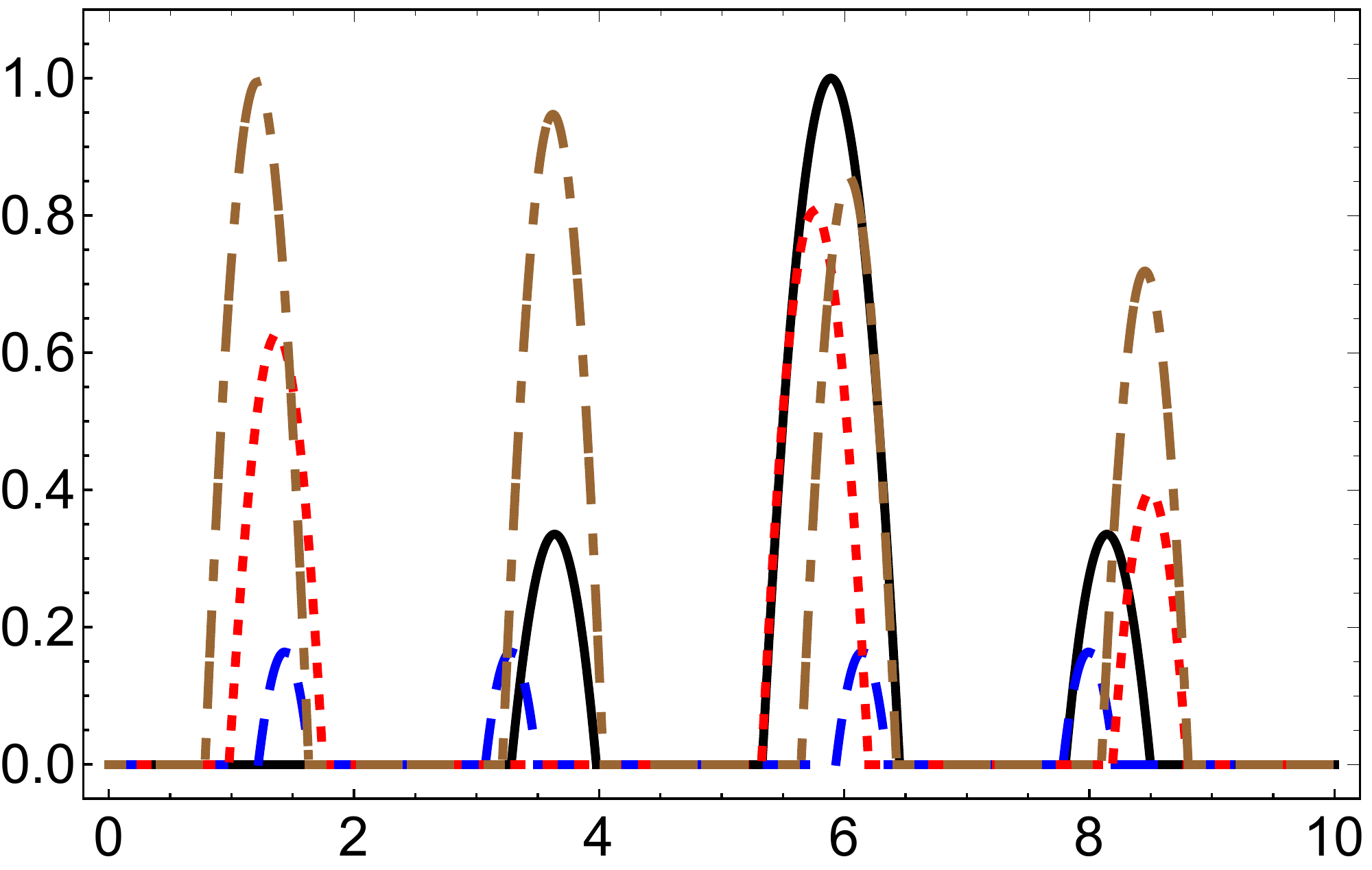}
		\put(-160,100){($ b $)}\put(-160,60){$\mathcal{N}_{La}$}
		\put(-70,-15){$\tau$}~~~\quad\quad
			\end{center}
	\caption{\label{dipolar7}(a) The behavior of the negativity $\mathcal{N}_{eg}$ and (b) The  non-local coherent advantage $\mathcal{N}_{La}$ of the channel  $ \rho_{23} $.  The solid, dash, the dot and the dash-dot lines represent the   $\mathcal{N}_{eg}$ and $\mathcal{N}_{La}$  at $\tilde{\epsilon}=-0.2, 0, 0.1, 0.3$, respectively. It is assumed that the initial network via two pairs of maximum entangled nodes ($MM$-type. }
\end{figure}
In Fig.(\ref{dipolar7}), we show the behavior of  both quantifiers at different values of the coupling constant $\tilde{\epsilon}$. It is clear that, the sudden death phenomenon  of both quantifiers  appears periodically. However, the sudden death time of the non-local coherent advantage is much larger than that displayed for the negativity.  The upper bounds of $\mathcal{N}_{eg}$ and $\mathcal{N}_{La}$ depend on the direction of the interaction, where when the interaction  is switched in $x-y$ plane, the negativity is much larger than the non-local coherent  advantage. On the other hand, the smallest  values of the non-local coherent advantage are displayed at zero coupling, namely, $\epsilon=0$, while $\Delta$ is arbitrary.

In this context, it is important to mention that, the behavior of the negativity and the non-local coherent advantage  between the second and the third nodes is similar for any initial entangled network.  Moreover, our calculations show that, there is no any quantum correlations are generated between the other nodes  as, the first and the third nodes  namely $\rho_{13}$, and between the second and fourth nodes $\rho_{24}$.  These encourage to discuss the possibility of finding quantum correlations between three different nodes. The next subsection is devoted for this aim, where we quantify the amount of entanglement  that may be generated between each three nodes by using the tangle.

\subsection{Three entangled nodes}
In this subsection, we investigate the possibility of generating  quantum correlation between any three different nodes via the spin Dipolar interaction. The amount of the quantum correlations are quantified by means of the tangle between all the three possible partitions, namely $\rho_{123}, \rho_{124}$ and $\rho_{234}$.  Our results show that, the behavior  of the tangle is similar for all the three  partitions. Therefore, we consider the behavior of $\mathcal{T}_{123}$ only  against the interaction strength $\tilde\epsilon$.
\begin{figure}[!h]
	\begin{center}
		\includegraphics[width=0.45\textwidth, height=130px]{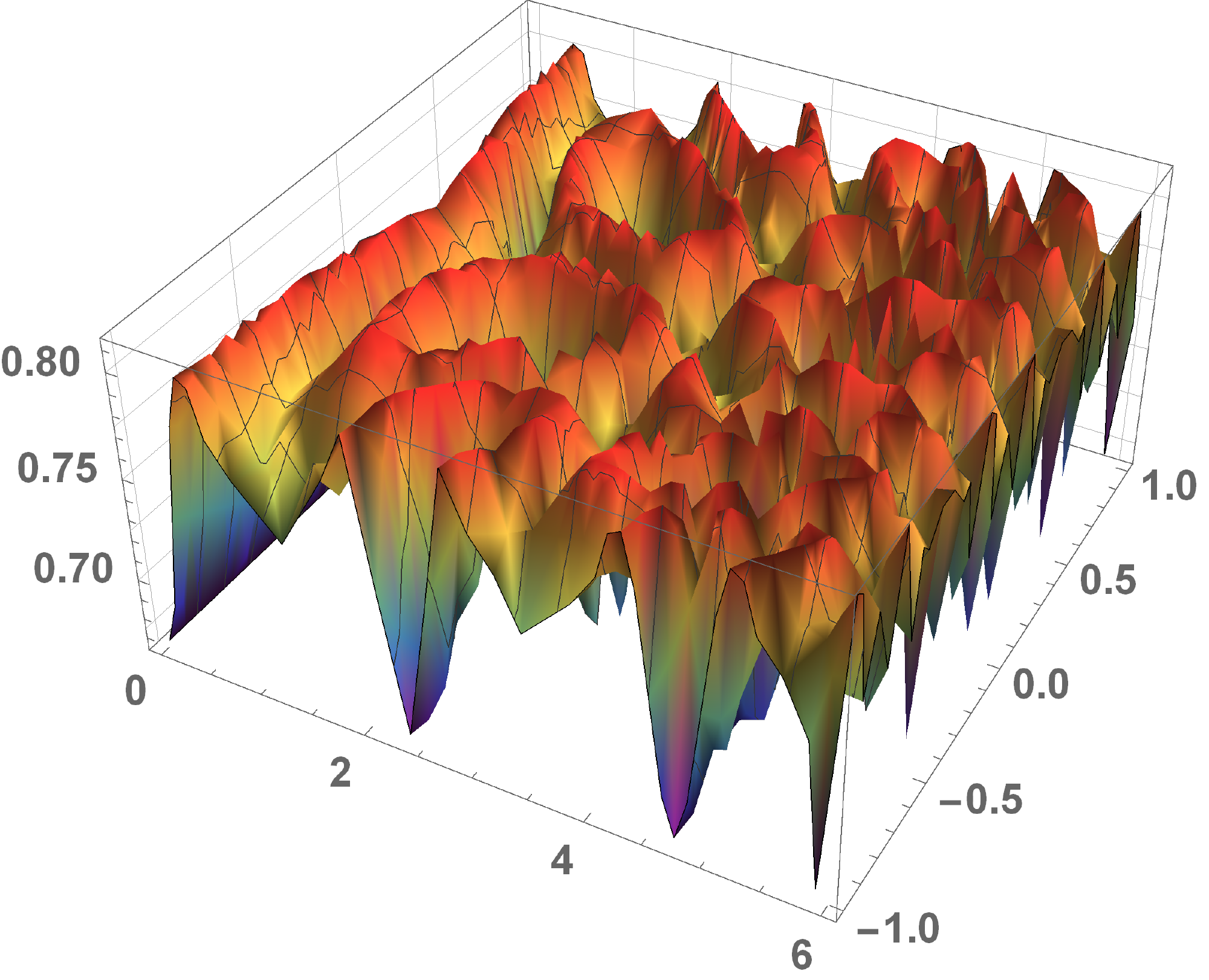}
		\put(-210,100){($ a $)}\put(-225,70){$\tau_{123}$}
		\put(-130,7){$\tau$}\put(-20,20){$\tilde{\epsilon}$}~~~~~~\quad\quad
\includegraphics[width=0.4\textwidth, height=130px]{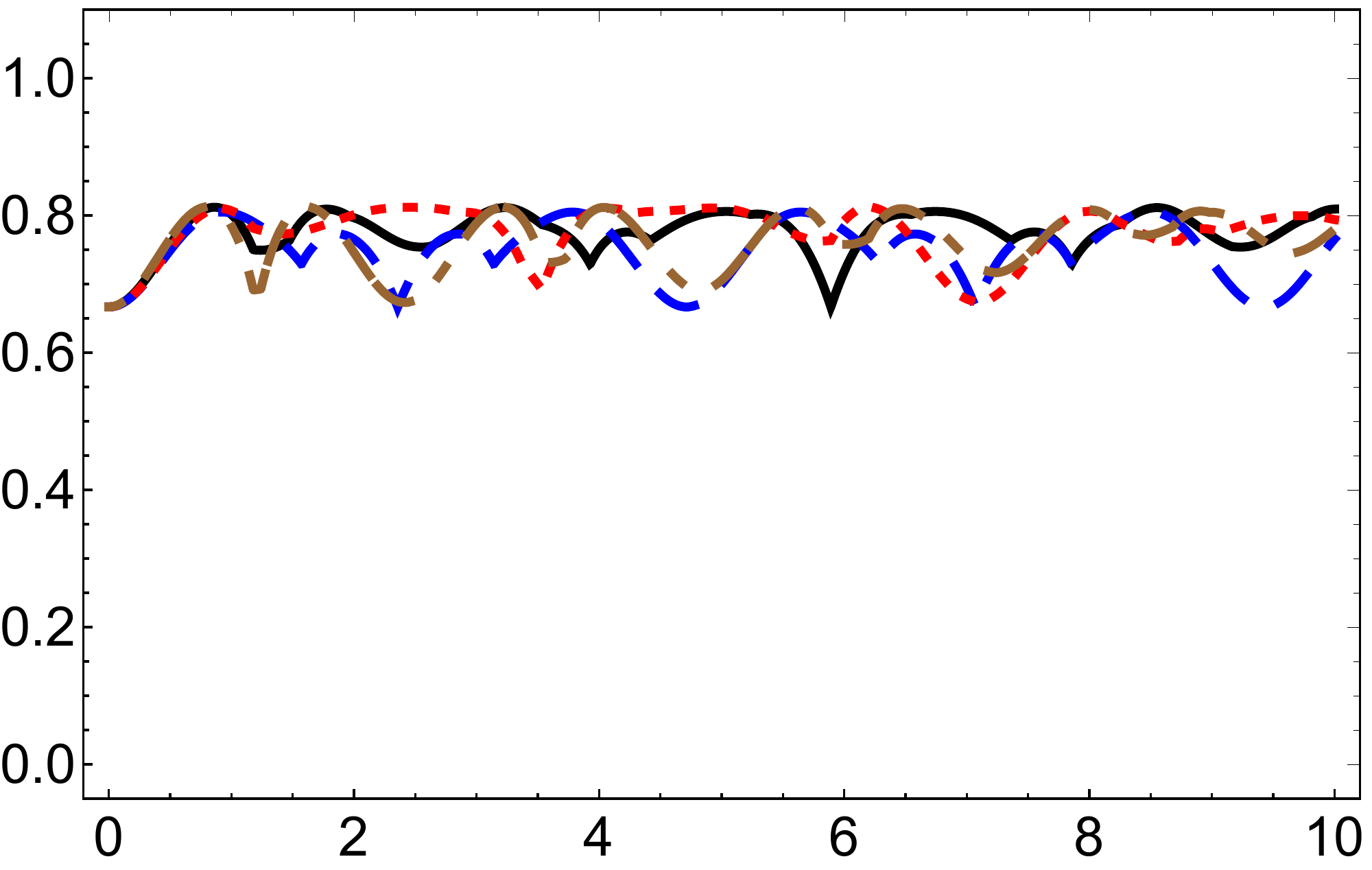}
		\put(-200,100){($ b $)}\put(-210,70){$\tau_{123}$}
		\put(-70,-15){$\tau$}~~~\quad\quad
	\end{center}
	\caption{\label{dipolar8}(a)The behavior of the tangle, $\tau_{123}(\tau,\tilde\epsilon)$ of $ \rho_{123} $, where the initial network of $MM$ type, and  (b)   represents  $\mathcal{T}_{123}(\tau)$, where the solid, dash, the dot and the dash-dot lines are evaluated at  $\tilde{\epsilon}=-0.2, 0, 0.1, 0.3$, respectively.}
\end{figure}
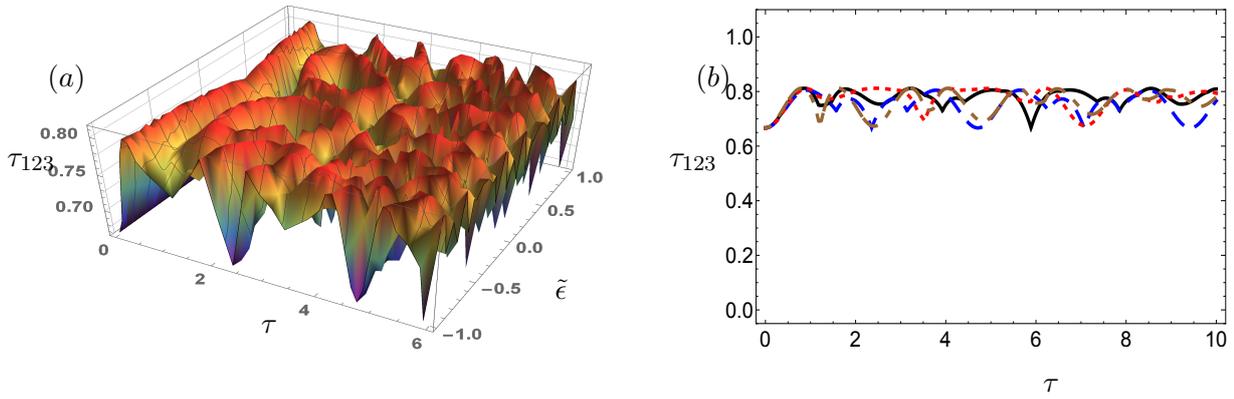

Fig.(\ref{dipolar8}), describes the behavior of the amount of entanglement that generated between the first, second and the third nodes by using the tangle.  As it  is displayed from Fig.(\ref{dipolar8}a), the tangle $\mathcal{T}_{123}(\tau,\tilde\epsilon)$ behaves as peaks, where it oscillates between its maximum and minimum bounds. Moreover, the phenomenon of the sudden death doesn't depicted. However, the sudden changes of the tangle are observed periodically, but never vanishes and its minimum values are larger than $0.65$.   The effect of some values of the interaction strength  is displayed in Fig.(\ref{dipolar8}b), where the amplitudes of oscillations are  small and consequently, the minimum values are much better than those displayed for the entangled two nodes. As it is shown from Fig.(\ref{dipolar8}b), the effect of interaction  strengths on the tangle is almost similar.

Comparing the  behavior of the tangle, which quantifies the amount of quantum correlation between any  three different nodes, and the negativity, which represents the entanglement between two nodes, one may  conclude that, although the negativity reaches its maximum values at some cases, but it suffers from sudden death during the interaction  time. However, the amount of quantum correlation between any  three nodes are more robust and never vanishes.  Moreover, the behavior of the tangle shows that, the upper bounds of quantum correlation between a thee qubits are much larger than that displayed between any two nodes. Also, the initial tangle depends on the initial entangled nodes, namely it is larger than zero at $\tau=0$.

\subsection{Extension network}
In this subsection, we extend our work to includes 8 nodes, where each 4 nodes conduct a hop. It is assumed that, the both terminals of the two hops interact locally   via the spin Dipolar interaction and consequently  an entangled network consists of $8$ nodes is generated. By tracking all the intermediate states, one obtains a final channel between the first and eighth nodes, $\rho_{18}$.

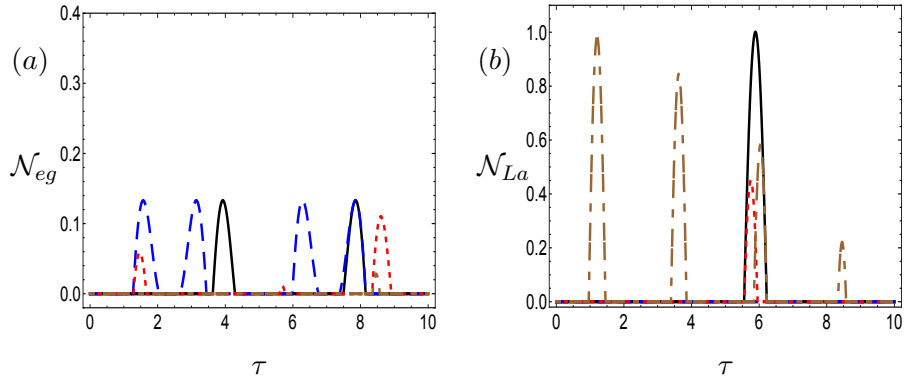
\begin{figure}[!h]
	\begin{center}
		\includegraphics[width=0.3\textwidth, height=125px]{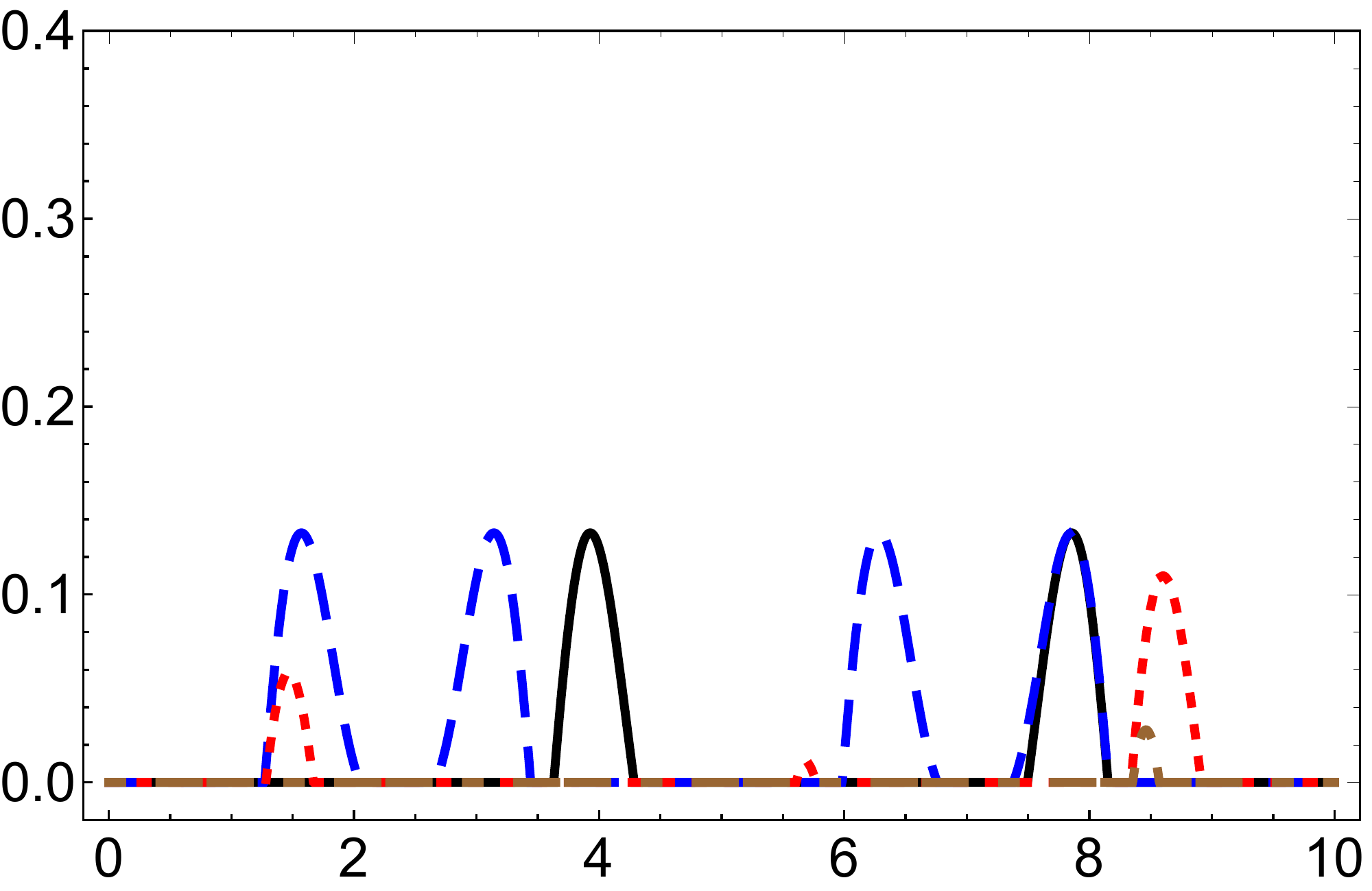}
		\put(-160,100){($ a $)}\put(-160,60){$\mathcal{N}_{eg}$}
		\put(-70,-15){$\tau$}~~~\quad\quad
		\includegraphics[width=0.3\textwidth, height=125px]{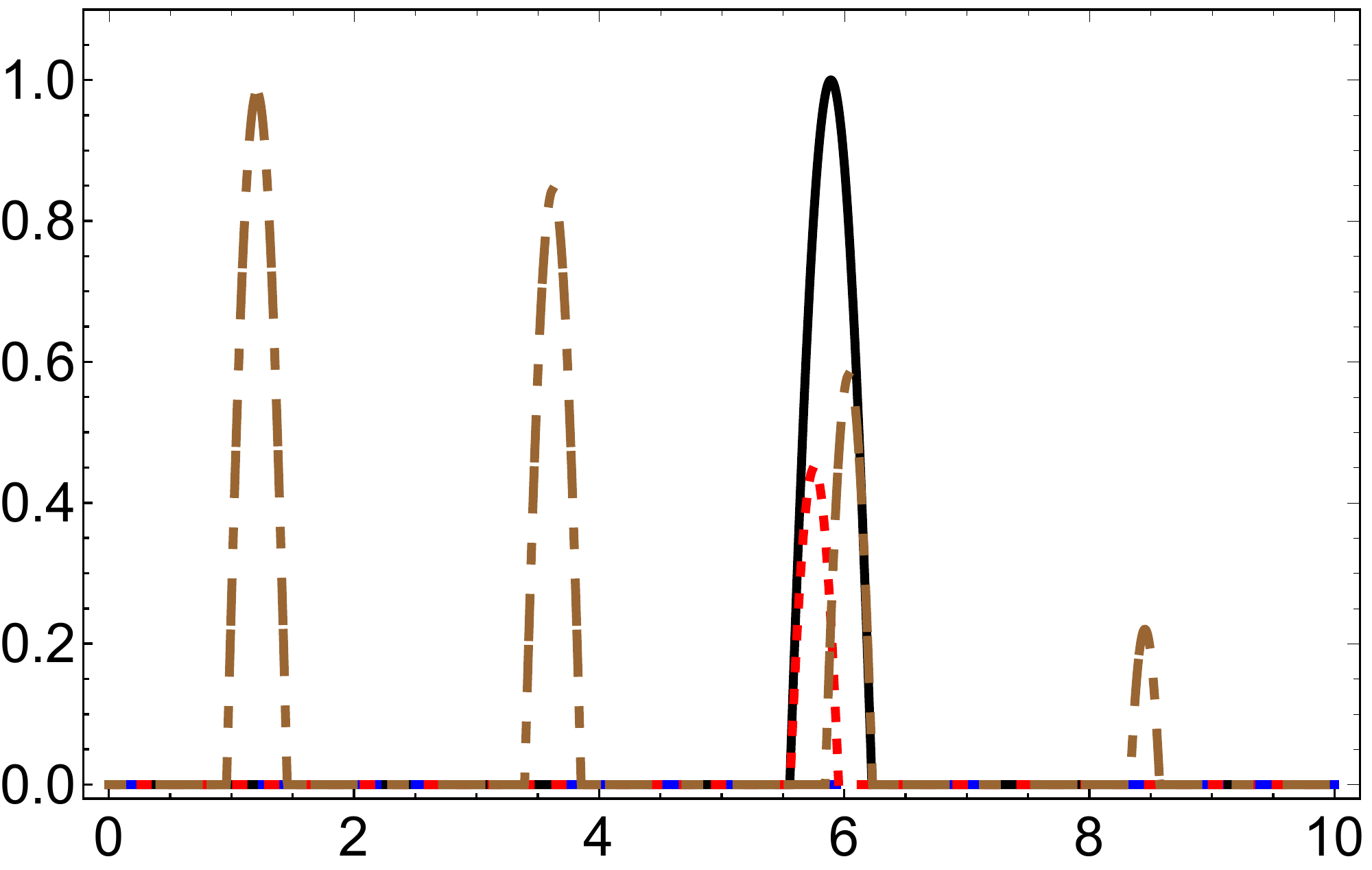}
		\put(-160,100){($ b $)}\put(-160,60){$\mathcal{N}_{La}$}
		\put(-70,-15){$\tau$}~~~\quad\quad
	\end{center}
	\caption{\label{dipolar9} The behavior of (a) the negativity $\mathcal{N}_{eg}$ and (b)the non- local coherent advantage, $\mathcal{N}_{La}$  of the channel $\rho_{18}$. The solid, dash, the dot and the dash-dot lines represent the   $\mathcal{N}_{eg}$ and $\mathcal{N}_{La}$  at  $\tilde{\epsilon}=-0.2, 0, 0.1, 0.3$, respectively, where the initial two hops are prepared via maximum entangled nodes ($MM$-type).  }
\end{figure}

The behavior of entanglement $\mathcal{N}_{eg}$ and the non local coherent advantage  $\mathcal{N}_{La}$  is displayed in Fig.(\ref{dipolar9}). The  non-zero values of both quantifiers are depicted when both hops are prepared initially  by using maximum entangled nodes.  The amount of entanglement is very small compared with those between $\rho_{14}$ and $\rho_{23}$ and $\rho_{12}$. Moreover, the upper bounds of the non-local coherent advantage are much larger than those displayed for the entanglement.  Switching  on the interaction in $z$- direction  predicted a large amount of $\mathcal{N}_{La}$, where it vanishes if it is switched in $x-y$ plane

\section{Conclusion}\label{Dip5}
The possibility of using the spin Dipolar interaction to generate quantum  entangled network by using multi-nodes is discussed. It is assumed that, each two nodes are initially prepared either in a  maximally or partially correlated, namely prepared in a singlet or Werner states, respectively. The suggested  network  is constructed by  four nodes, where the connection either direcet/inderict via spin  Dipolar interaction. Moreover,  it is allowed that, the terminals of the  two hops interact directly, to generate an entangled hop consists of   eight nodes. By using the same technique, this network  may be extended to multi-hops  entangled network.

Due to the interaction, there is a  quantum(classical) correlation  is generated between each   two or three nodes. The predictable  quantum correlations are  investigated via nonlocal coherent advantage and quantified by the negativity for each  entangled two nodes and by the tangle for the entangled three nodes. The effect of the interaction strength and the  initial entangled network  on the generated network  is discussed. Our results show that, due to the symmetric, each initial entangled nodes behave  similarly under the effect of the interaction strength, namely the quantum correlations between the first and the second nodes,  is  the same  as that  between the third and the fourth nodes. The  predicted amount of quantum correlations  between each  three nodes is affected   in a  similar way  by the interaction strength. Moreover, there is only classical correlations are generated between the first-third and the second-fourth nodes.

 It is shown that,  the phenomena of the sudden death/birth is displayed for all  the entangled two  nodes, either generated via direct or indirect  interaction. However, for the entangled three nodes, the phenomena of the sudden changes (increasing or decreasing) is displayed.  The number of peaks  of the negativity and the  non-local coherent advantage that  has  predicted between all the nodes increases as the interaction increase.  Starting from the $MM$ initial network, the upper bounds that predicted is much larger than those displayed if the initial nodes are prepared in a partial entangled nodes.

However, the behavior of the negativity and the non-local coherent advantage depends on the connection method, initially entangled, directly or non-directly interaction. It is shown that, for initially maximum  entangled nodes network, the upper bounds of the negativity and the non-local coherent advantage reach their maximum values periodically, depending on the interaction strength.  However, when the interaction is switched in $z-$ direction with large interaction strength,  the number of oscillations of both quantifiers is larger than those displayed in $x-y$ plane. Moreover, at zero coupling, the oscillations that displayed for the negativity and the non-local coherent advantage have maximum values and large death time. The survival behavior of both measures, decreases when the network is initially constructed from  partially entangled nodes, ($WW$ network). For the indirect interaction nodes, the behavior of both phenomena is similar to those displayed for initially entangled  nodes, but the oscillations  are faster and their maximum bounds are smaller. The smallest upper bounds are predicted when the initial entangled nodes are prepared in  partially entangled nodes, ($WW$ network). On the other hand,  for the direct connected nodes, the entanglement  never reaches its maximum values even the initial network is prepared via maximum entangled nodes.

The behavior of the generated quantum  correlations between  each   three nodes  is examined, where we assume that the initial network is of type $MM$ network. It is shown that, the amount of correlations between each three qubits is almost  equalize. The sudden changes (decreasing/ increasing) of the tangle are  displayed during the interaction time.  The amplitudes  of the tangle   oscillations are smaller and consequently the minimum values are improved.  The  direction of the  interaction strength almost similar, where the oscillations of the tangle has the same  amplitudes.

The possibility of extending the network into multi-hops network is  examined, where we quantify the amount of entanglement and the non-local coherent advantage that are  generated between the initial terminal of the first hop and the last node on the second hop. It is shown that, the amount of correlation decreases and the time of the sudden death   increases.

{\it In conclusion}, the spin Dipolar interaction could be used to generated quantum correlations either between two or three nodes via direct or indirect interaction. The  amount of the generated correlations depend on the initial network settings. The interaction strength and its direction has a remarkable effect on the  behavior  of the generated correlations between each two nodes, while  there is a slightly effect of that  generated between the three qubits.

‏	
\end{document}